\documentclass{article}
\usepackage[utf8x]{inputenc}
\usepackage[american]{babel}  
\usepackage{graphicx}
\usepackage[mathcal]{euscript}
\usepackage{mathrsfs}
\usepackage{framed}
\usepackage{amssymb}
\usepackage{nopageno}
\usepackage{amsmath}
\usepackage{amsfonts}
\usepackage{esint}
\usepackage{natbib}  
\usepackage{caption}
\usepackage{subcaption}
\usepackage[colorlinks=true,citecolor=blue]{hyperref}
\title{Mixed-Membership of Experts Stochastic Blockmodel}
\author{Arthur White \footnote{School of Computer Science and Statistics, Trinity College Dublin, Dublin 2, Ireland.} and Thomas Brendan Murphy\footnote{ School of Mathematical Sciences, University College Dublin, Dublin 4,
Ireland.}~\footnote{This work is supported by Science Foundation Ireland under the Clique Strategic Research Cluster
(SFI/08/SRC/I1407) and the Insight Research Centre (SFI/12/RC/2289).
}}
\newcommand{\Esub}[2]{{\mathbb E}_{#2}\left[{#1}\right]}
\newcommand{\Binom}[2]{{#1}^{#2}(1- #1)^{1 - #2}}

\newcommand{\del}[2]{\frac{\partial #2}{\partial #1}}
\newcommand{\deltwo}[3]{\frac{\partial^2 #3}{\partial #1 \partial #2}}

\begin{document}
\date{}
\maketitle

\begin{abstract}
 Social network analysis is the study of how links between a set of actors are formed. Typically, it is believed that links are formed in a structured manner, which may be due to, for example, political or material incentives, and which often may not be directly observable. The stochastic blockmodel represents this structure using latent groups which exhibit different connective properties, so that conditional on the group membership of two actors, the probability of a link being formed between them is represented by a connectivity matrix. The mixed membership stochastic blockmodel (MMSBM) extends this model to allow actors membership to different groups, depending on the interaction in question, providing further flexibility.

Attribute information can also play an important role in explaining network formation. Network models which do not explicitly incorporate covariate information require the analyst to compare fitted network models to additional attributes in a post-hoc manner. We introduce the mixed membership of experts stochastic blockmodel, an extension to the MMSBM which incorporates covariate actor information into the existing model. The method is illustrated with application to the Lazega Lawyers dataset. Model and variable selection methods are also discussed.
\end{abstract}

\section{Introduction}
Social network analysis (SNA)~\citep{Wasserman1994,airoldi2007,salter12} is the study of how links between a set of actors are formed. Typically, it is believed that links are formed in a structured manner, so that the Erd\H{o}s-R\'{e}nyi model~\citep{Erdos1959}, whereby links occur independently with a constant probability throughout the network, fails to capture many aspects of real-world datasets. Reasons for this structure may be due to, for example, political or material incentives, and often may not be directly observed.

Several classes of statistical methods have been proposed to examine this structure. Exponential family random graph models \citep{Holland1981,Snijders2002,Robins2006} examine whether subgraph summary statistics occur significantly more frequently than by random chance in an unstructured network. If this is the case, then this is treated as evidence of a particular underlying mechanism in the network structure. For example, a larger number of triangles than could reasonably be expected by chance occurring in a network is evidence of transitivity, whereby a mutually shared link to an actor increases the probability of a link between two actors.

Two other approaches represent network structure using latent variables. The stochastic blockmodel (SBM)~\citep{holland1983,Snijders1997,Daudin2008} introduces $G$ latent groups underlying the network, so that conditional on the group membership of two actors, the probability of a link being formed between them is represented by a $G \times G$ connectivity matrix. The latent space network model~\citep{Hoff2002} 
 maps actors onto a $d$-dimensional space so that the probability of a link being formed between two actors becomes a function of their distance from each other. The latent position cluster model \citep{Handcock2007} then extends this model so that the positions of actors in this space are determined by a mixture of spherical multivariate normal distributions. Both the SBM and latent position cluster models can be thought of as types of mixture model applied to network data.
 
A key difference between the models is that the latent space model is constrained to cluster together actors with strong connections with each other but weak connections to other actors in the network, a behaviour known as \emph{affiliation}. Conversely, the SBM has no such constraints and can represent this behaviour, as well as \emph{disassociative mixing}, whereby disparate actors connect strongly to a distinct set of actors but only weakly with each other \citep{latouche2011}. \citet{airoldi2008} and \citet{latouche2011} develop extensions to the SBM, introducing mixed-membership (MMSBM) and overlapping SBMs respectively. These models allow actors membership of different groups, depending on the actors with which they are interacting, further extending the flexibility of the SBM. 

Attribute information can also play an important role in helping to explain how a particular network structure has occurred. For example, high school students might be more likely to form friendships with others in the same class as them, while gender plays an important role in  the formation of sexual networks. This belief, referred to as ``homophilly by attributes'' is reflected by ~\cite{Breiger1974}, who notes the ``metaphor which has often appeared in sociological literature,'' that ``groups \ldots are collectivities based on the shared interests, personal affinities, or ascribed status of members who participate regularly in collective activities.''


Network models which do not incorporate covariate information require the analyst to compare fitted network model clusterings to additional attributes in a post-hoc manner  \citep{Handcock2007,airoldi2008}. \citet{mariadassou2010} and \citet{Gormley2010} respectively extend the SBM and latent space models to incorporate covariates, at link and actor-specific levels. Examples of actor-specific covariates include gender and age, while link-specific covariates relate additional information about the relationship between actors, such as the physical distance between actor locations. \citet{mariadassou2010} also introduce specifications for SBMs fitting for types of interaction beyond binary link types. These models can explicitly investigate the impact which concomitant covariate information has on network structure.

In this paper we present a method which incorporates actor attribute information into the MMSBM, the mixed-membership of experts stochastic blockmodel (MMESBM). This method makes use of the mixture of experts terminology framework introduced by \citet{jacobs1991} to allow model parameters to depend on covariate information; we adapt the terminology since the covariates are incorporating into a mixed-membership rather than mixture model framework. This model may be thought of as a type of social selection model \citep[e.g.,][]{fellows2012}, in that it is assumed that actor characteristics influence network formation, while the attribute information itself is assumed fixed and known. Models where the converse applies, so that social ties are seen as influencing actor characteristics, are referred to as social influence models.

The rest of the paper is structured as follows: both the SBM and MMSBM are briefly reviewed, before the MMESBM is introduced in Section~\ref{sec:modelspec}. A variational Bayes method for inference similar to that proposed by \citet{airoldi2008} is then described in Section~\ref{sec:inference}. Model selection and validation methods are also discussed in this section. The model is applied to the Lazega Lawyers dataset in Section~\ref{sec:results}. The results are interpreted and some goodness fit diagnostics are also performed. Possible further extensions to the model are then discussed in Section~\ref{sec:conclusion}. Some additional details on model inference are provided in Appendix~\ref{sec:covar}.

\section{Model Specification}\label{sec:modelspec}
Relational data consists of a set of actors $a_1, \ldots, a_N$, and the links which they share with each other. In this paper we assume that the links are binary valued, i.e., that they are present or absent. Let the adjacency matrix $\mathbf{Y}$ represent the interaction between pairs of actors in a network. An interaction  between any pair of actors $a_i$ and $a_j$ can then be represented as
\begin{equation*} Y_{ij} = \left\{ \begin{array}{ll}
1 & \mbox{if a link exists between actors $a_i$ and $a_j$;} \\
0 & \mbox{otherwise}.
\end{array} \right. 
\end{equation*} 

If the link type is thought of as being shared, or symmetric, then the network is said to be undirected, with $Y_{ij} = Y_{ji}$. Otherwise, it is said to be directed. In some settings, such as protein-protein interactions, self-interaction is possible, i.e., $Y_{ii}$ can take values. This property is referred to as reflexivity. In other cases, such as when friendship between high school students is being considered, such an interaction is not considered meaningful, making the network irreflexive, and as such the diagonal entries of $\mathbf{Y}$ are considered undefined. For the purposes of this paper, we consider only the case when a network is directed and irreflexive.

\subsection{Stochastic Blockmodel}
The SBM assumes that $G$ latent groups underly the data. Conditional on their memberships to groups $g$ and $h$ respectively, the interaction between two actors $a_i$ and $a_j$ is then modelled by a $G \times G$ interaction matrix $\boldsymbol{\Theta}$, such that ${\mathbb{P}}(Y_{ij} = 1) = \Theta_{gh}$. Let $\boldsymbol{\tau}$ denote the mixing proportions of the groups, so that ${\mathbb{P}}(\mbox{Group } g) = \tau_{g}$. Each actor $a_i$ is assigned a group membership indicator $\mathbf{Z}_i$, such that \begin{equation*} Z_{ig} = \left\{ \begin{array}{ll}
1 & \mbox{if actor $a_i$ belongs to Group $g$;} \\
0 & \mbox{otherwise}.
\end{array} \right. 
\end{equation*} 
Each $\mathbf{Z}_i$ then follows a multinomial distribution, with one trial and probability vector $\boldsymbol{\tau}$. The choice of conjugate priors ensures that $\boldsymbol{\Theta}$ and $\boldsymbol{\tau}$ follow beta and Dirichlet distributions respectively \citep{Snijders1997}, or inference can be performed in a frequentist framework \citep{Daudin2008}. Inference for the SBM is possible using a variational approximation \citep{Daudin2008} or a collapsed Gibbs sampler \citep{mcdaid12}. Gibbs sampling on the fully parameterised SBM is also possible \citep{Snijders2001}, although at substantial additional computational cost.

\subsection{Mixed-Membership Stochastic Blockmodel}
The MMSBM \citep{airoldi2008} extends the SBM to allow actors membership to multiple groups depending on the actor with which they interact. Within this framework, each actor $a_i$ is assigned an individual mixing parameter $\boldsymbol{\tau}_i$, denoting their propensity for group membership. Indicator vectors $\mathbf{Z}^{1}_{ij}$ and $\mathbf{Z}^{2}_{ij}$ (note the superscript indices) denote the group membership of actors $a_i$ (sender) and $a_j$ (receiver) during an interaction $Y_{ij}$. Conditional on this additional model complexity, actor interaction is again modelled by a matrix $\boldsymbol{\Theta}$ in a similar manner to the SBM.\footnote{\cite{airoldi2008} also introduce an additional sparsity parameter in order to distinguish between the case where interactions in the network are in general quite rare, and when non-interaction is due to particularly low-level connection between groups. We exclude this parameter from our analysis.} Choosing a Dirichlet prior distribution with hyperparameter $\boldsymbol{\delta}$ ensures that each mixing parameter $\boldsymbol{\tau}_i$ also follows the same distribution. A beta distribution can also be specified for $\boldsymbol{\Theta}$ with the choice of a conjugate prior, otherwise it may be treated as a nuisance parameter \citep{airoldi2008}.

\subsection{Mixed-Membership of Experts Stochastic Blockmodel}
The MMSBM can be further extended by allowing the parameters of the model to be functions of concomitant covariate data. The terminology used in the mixture of experts literature refer to functions of covariates and mixing parameters  as ``gating networks''\footnote{Note that in this terminology, the network in question refers to the graphical model specification, and is not to be confused with the network data under investigation.} and functions of covariates and conditionally distributive parameters as ``experts'' \citep{Gormley2010}. In this paper we restrict our analysis to actor-specific attributes $\mathbf{W}_i = W_{i1}, \ldots, W_{iP}$, 
which are incorporated into the prior distribution of the individual-level mixing parameters $\boldsymbol{\tau}$. The hyper-parameter $\boldsymbol{\delta}_i$ 
 is treated as a function of $\mathbf{W}$ and parameter $\boldsymbol{\beta}_g = \beta_{g1}, \ldots, \beta_{gP} $, $g= 1, \ldots, G$ such that $\delta_{ig}(\mathbf{W}_i) = \exp(\sum^P_{p=1} W_{ip}\beta_{gp}).$ Note that inference when including link-specific attributes in a mixed-membership setting may be treated in a similar fashion to the mixture framework described in \cite{mariadassou2010}. The data generative process for the MMESBM is outlined in Figure~\ref{fig:DataGen}.

%
%
%
%

%
%
%

\begin{figure}
\begin{framed}
\begin{itemize}
\item for $i  \in 1, \ldots, N$:

$\boldsymbol{\tau}_i \sim$ Dirichlet($\exp( \mathbf{W}^\top_{i}\boldsymbol{\beta})$).

\item for $g \mbox{ and } h \in 1, \ldots, G$:

$\boldsymbol{\Theta}_{gh} \sim$ Beta($\alpha_{gh}, \beta_{gh}$).
\end{itemize}
\begin{itemize}

\item for $i \mbox{ and } j \in 1, \ldots, N$:

$\mathbf{Z}^{1}_{ij} \sim $ Multinomial (1, $\boldsymbol{\tau}_{i}$),

$ \mathbf{Z}^{2}_{ij} \sim $ Multinomial (1, $\boldsymbol{\tau}_{j}$),


$Y_{ij} \sim$ Bernoulli($\mathbf{Z}^{1}_{ij}\boldsymbol{\Theta} \mathbf{Z}{^{2\top}_{ij}}$).
\end{itemize} 
\end{framed}
\caption{Data generative process for the MMESBM.}
\label{fig:DataGen}
\end{figure}

\begin{figure}[htbp]
\begin{center}
 \begin{subfigure}[b]{0.49\textwidth}
 \includegraphics[width=0.95\linewidth]{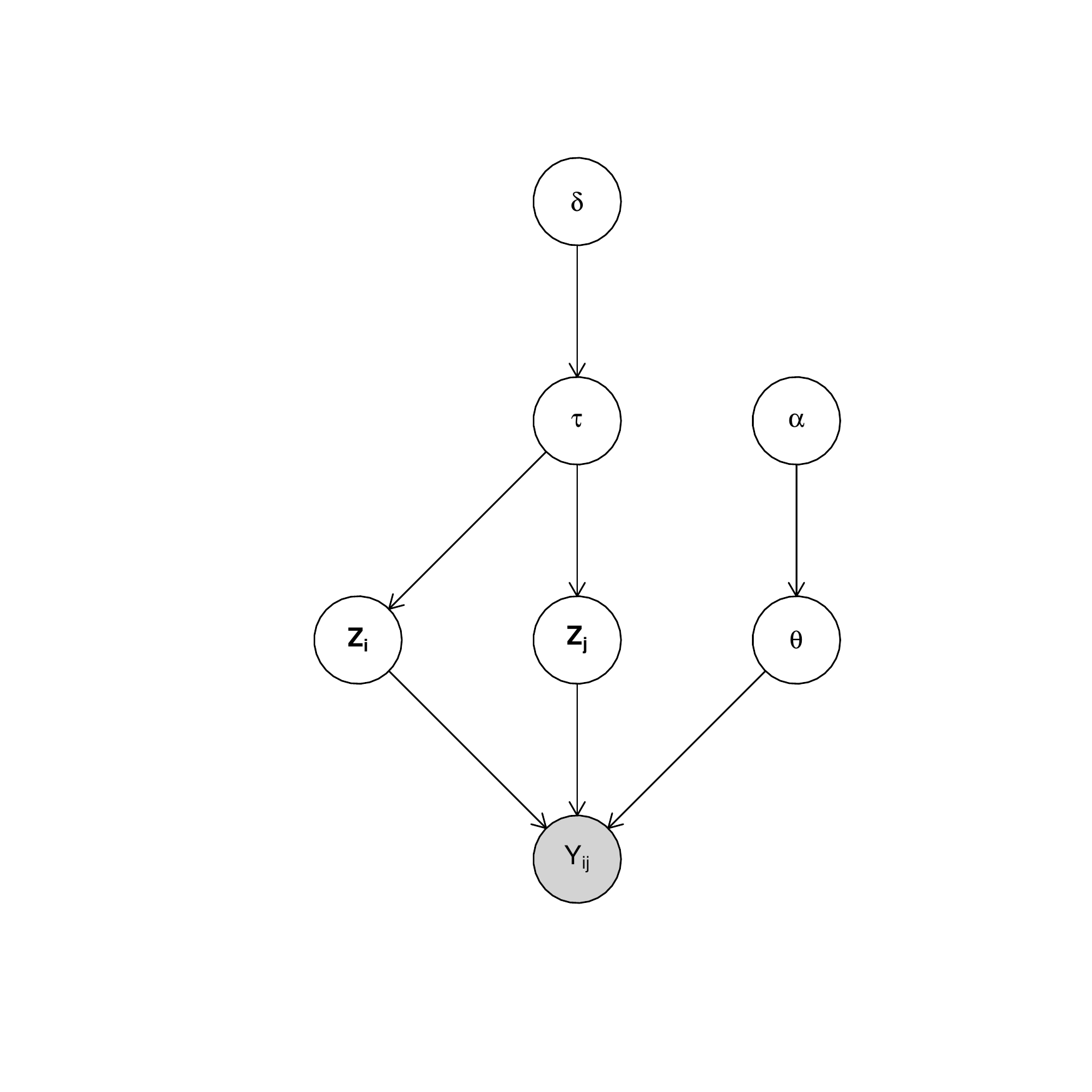}
               \caption{Stochastic blockmodel}
                \label{fig:GraphPlota}
         \end{subfigure}
 \begin{subfigure}[b]{0.49\textwidth}
 \includegraphics[width=0.95\linewidth]{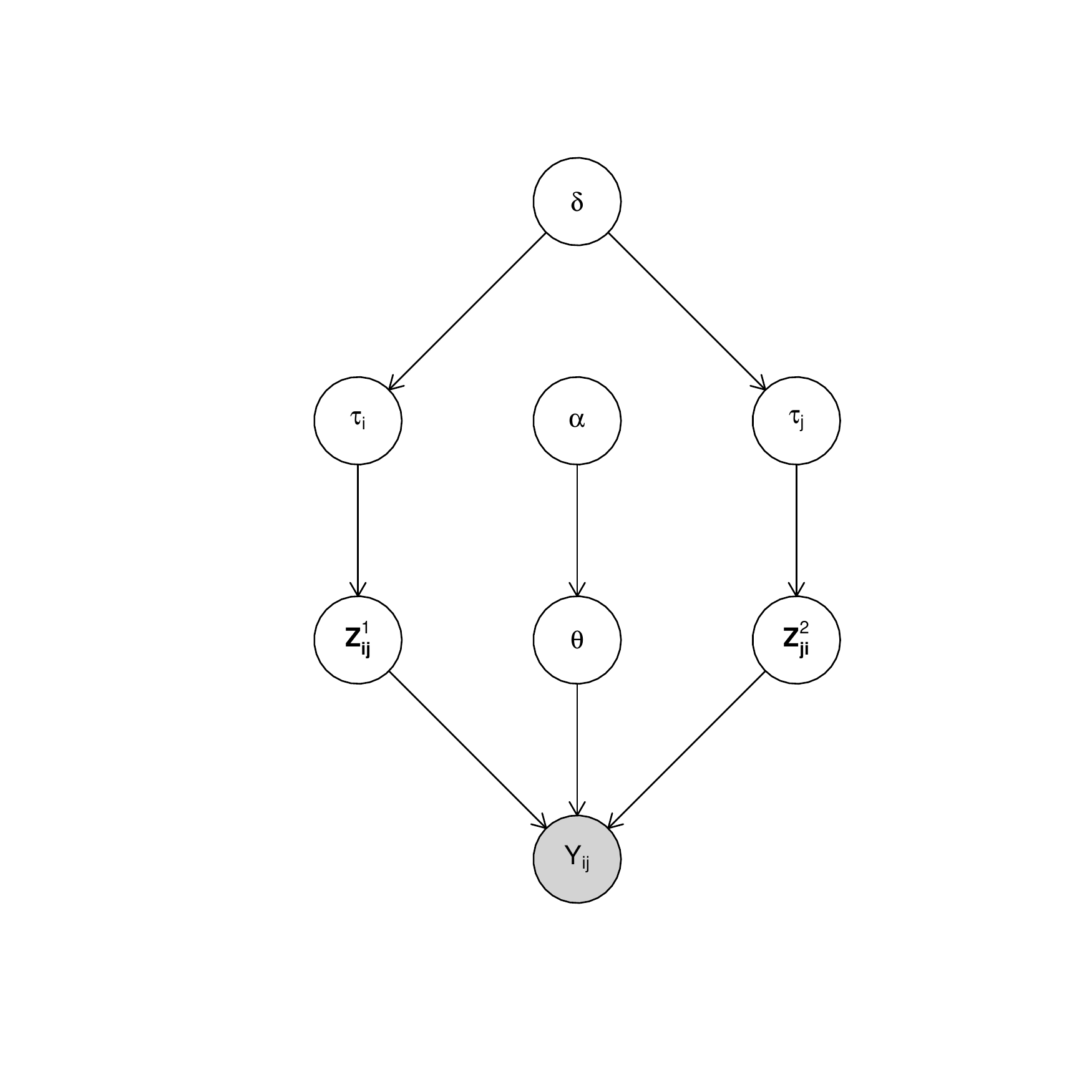}
               \caption{Mixed-membership stochastic blockmodel}
                \label{fig:GraphPlotb}
         \end{subfigure}
 \begin{subfigure}[b]{0.49\textwidth}
 \includegraphics[width=0.95\linewidth]{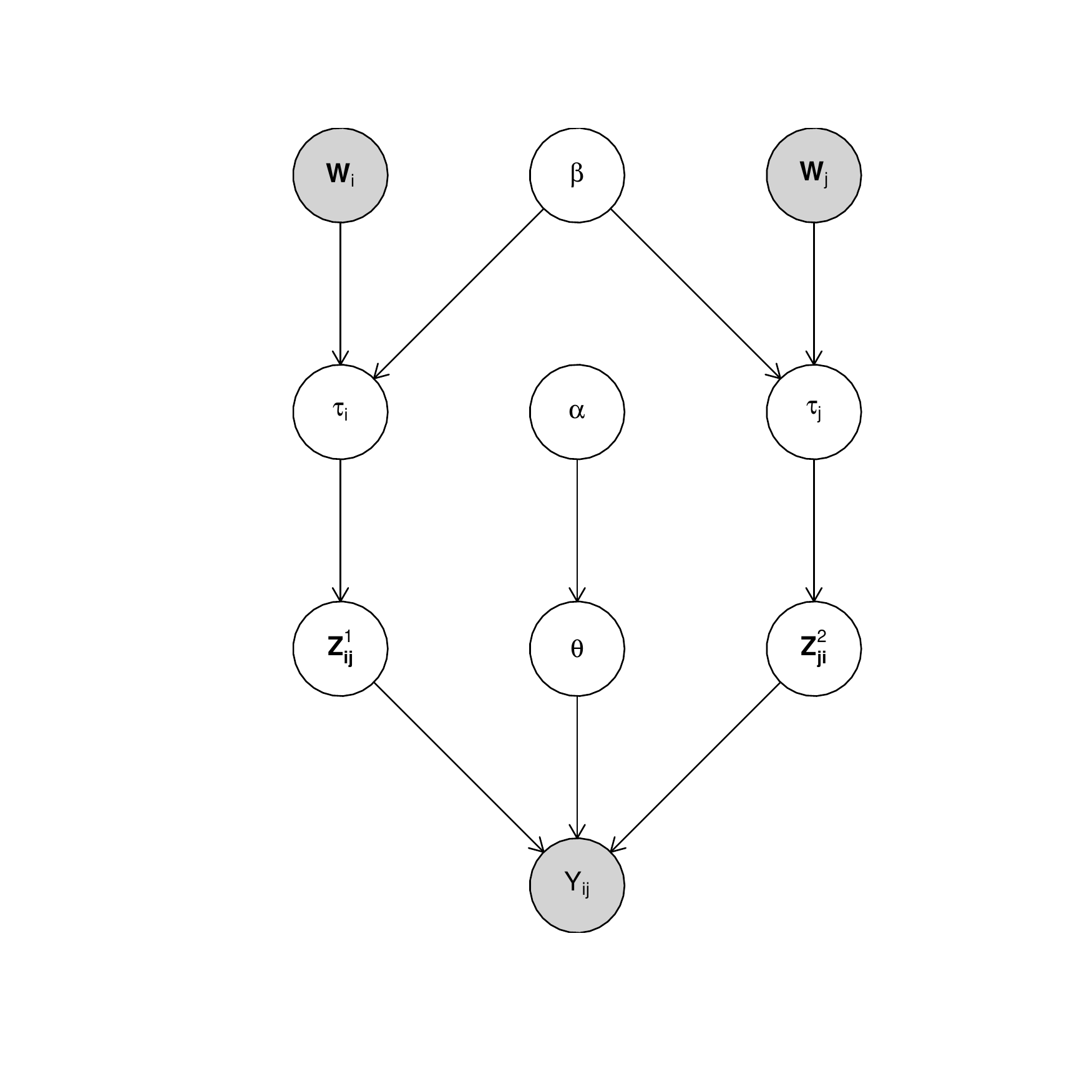}
                \caption{Mixed-membership of experts stochastic blockmodel}
                \label{fig:GraphPlotc}
         \end{subfigure}
\caption{Graphical model representations of the SBM (a), MMSBM (b) and MMESBM (b). Note the different position of $\Theta$ in (a), chosen for graphical simplicity. }
\end{center}
\end{figure}

The model posterior can be decomposed thus:
\begin{equation}
 p(\mathbf{Y}, \mathbf{Z}^1, \mathbf{Z}^2, \boldsymbol{\tau}, \boldsymbol{\Theta} | \boldsymbol{\alpha}^1, \boldsymbol{\alpha}^2,  \boldsymbol{\beta}, \mathbf{W}) = \prod^N_{i=1} \prod^N_{j=1, j\neq i} p(Y_{ij} | \mathbf{Z}^1_{ij}, \mathbf{Z}^2_{ij}, \boldsymbol{\Theta})p(\mathbf{Z}^1_{ij} | \boldsymbol{\tau_i})p(\mathbf{Z}^2_{ij} | \boldsymbol{\tau_j}) \prod^N_{n=1} p(\boldsymbol{\tau}_{n} | \boldsymbol{\beta}, \mathbf{W}) p(\boldsymbol{\Theta} | \boldsymbol{\alpha}^1, \boldsymbol{\alpha}^2),\label{eq:ModelPost}
\end{equation}

where

\begin{eqnarray*}
p(Y_{ij} | \mathbf{Z}^1_{ij}, \mathbf{Z}^2_{ij}, \boldsymbol{\Theta}) &=&  \prod^G_{g=1} \prod^G_{h=1} \left \lbrace \Binom{\Theta_{gh}}{ Y_{ij}} \right \rbrace^{Z^{1}_{ijg}Z^{2}_{ijh}}\\
p(\mathbf{Z}^1_{ij} | \boldsymbol{\tau}_i) &=& \prod^G_{g=1} \tau_{ig}^{Z^1_{ijg}} \\
p(\mathbf{Z}^2_{ij} | \boldsymbol{\tau}_j) &=& \prod^G_{g=1} \tau_{jg}^{Z^2_{ijg}} \\
p(\boldsymbol{\tau}_n | \boldsymbol{\beta}, \mathbf{W}) &=& \frac{\Gamma \left(\sum^G_{h=1} \exp \left( \sum^P_{p=1} W_{np} \beta_{ph}\right) \right)}{\prod^G_{h=1}\Gamma \left ( \exp \left( \sum^P_{p=1} W_{np} \beta_{ph} \right) \right) } \prod^G_{g=1} \tau_{ng}^{\exp \left ( \sum^P_{p=1} W_{np} \beta_{pg} \right) -1} \\
p(\Theta | \boldsymbol{\alpha}^1, \boldsymbol{\alpha}^2) &=& \prod^G_{g=1} \prod^G_{h=1} \frac{\Gamma(\alpha^1_{gh} + \alpha^2_{gh})}{\Gamma(\alpha^1_{gh})\Gamma(\alpha^2_{gh})} \Theta_{gh}^{ \alpha^1_{gh} -1}(1-\Theta_{gh})^{ \alpha^2_{gh} -1}.
\end{eqnarray*}
Note that we again use superscript indices for the hyperparameters $\boldsymbol{\alpha}^1 \mbox{ and } \boldsymbol{\alpha}^2$. In what follows in Section~\ref{sec:results} we set ${\alpha}_{gh}^1 = {\alpha}_{gh}^2 = 1$ for $g,h = 1, \ldots, G$.
Graphical model representations of the SBM, MMSBM and MMESBM are provided in Figures \ref{fig:GraphPlota}--\ref{fig:GraphPlotc}.

\section{Model Inference}\label{sec:inference}
In a similar fashion to \citet{airoldi2008}, we estimate model parameters by employing a variational Bayes approximation. These have previously proved useful in both network \citep{Daudin2008,SalterTownshend2013} and mixed-membership settings \citep{blei03,Rogers05,erosheva07}. See \citet{beal03}, \citet[Chapter 10,][]{bishop2006} and \citet{Ormerod10} for overviews of the method at differing levels of intensity. The main idea is to approximate the posterior  $p(\mathbf{Z}^1, \mathbf{Z}^2, \boldsymbol{\tau}, \boldsymbol{\Theta})$ with a set of distributions $q(\mathbf{Z}^1, \mathbf{Z}^2, \boldsymbol{\tau}, \boldsymbol{\Theta})$ which have a nice form. 
Then the marginal log posterior can be re-written as: 
\begin{eqnarray*}
\log p(\mathbf{Y} | \boldsymbol{\beta}, \mathbf{W}) &=& \log \int_{\boldsymbol{\theta}} \int_{\boldsymbol{\tau}} \sum_{\mathbf{Z}^1} \sum_{\mathbf{Z}^2} p(\mathbf{Y}, \mathbf{Z}^1, \mathbf{Z}^2, \boldsymbol{\tau}, \boldsymbol{\Theta} | \boldsymbol{\alpha}, \boldsymbol{\beta}, \boldsymbol{\delta}) \frac{q(\mathbf{Z}^1, \mathbf{Z}^2, \boldsymbol{\tau}, \boldsymbol{\Theta})}{q(\mathbf{Z}^1, \mathbf{Z}^2, \boldsymbol{\tau}, \boldsymbol{\Theta})} d\boldsymbol{\tau} d\boldsymbol{\theta} \\
&\geq&  \int_{\boldsymbol{\theta}}  \int_{\boldsymbol{\tau}} \sum_{\mathbf{Z}^1} \sum_{\mathbf{Z}^2} q(\mathbf{Z}^1, \mathbf{Z}^2, \boldsymbol{\tau}, \boldsymbol{\Theta}) \log \frac{ p(\mathbf{Y}, \mathbf{Z}^1, \mathbf{Z}^2, \boldsymbol{\tau}, \boldsymbol{\Theta} | \boldsymbol{\alpha}, \boldsymbol{\beta}, \boldsymbol{\delta})}{q(\mathbf{Z}^1, \mathbf{Z}^2, \boldsymbol{\tau}, \boldsymbol{\Theta})} d\boldsymbol{\tau} d\boldsymbol{\theta}, \\
 &=& \Esub{ \log p(\mathbf{Y}, \mathbf{Z}^1, \mathbf{Z}^2, \boldsymbol{\tau}, \boldsymbol{\Theta} | \boldsymbol{\alpha}, \boldsymbol{\beta}, \boldsymbol{\delta})}{\mathbf{Z}^1, \mathbf{Z}^2, \boldsymbol{\tau}, \boldsymbol{\Theta}} - \Esub{\log q(\mathbf{Z}^1, \mathbf{Z}^2, \boldsymbol{\tau}, \boldsymbol{\Theta})}{\mathbf{Z}^1, \mathbf{Z}^2, \boldsymbol{\tau}, \boldsymbol{\Theta}},\\
 &=& {\cal L}.
\end{eqnarray*}
Here the concavity of the logarithmic function has been exploited to ensure that ${\cal L}$ is a lower bound to $\log p(\mathbf{Y} | \boldsymbol{\beta}, \mathbf{W})$, with the discrepancy in the inequality being equal to the Kullback-Liebler divergence \citep{Kullback1951} ${\cal KL}(q || p)$ between the true and approximate distributions $p$ and $q$. 

If we then restrict the set of distributions $q$ such that they can be factorized independently, then the optimal (i.e. the Kullback-Liebler divergence minimising) form of each distribution will be the same as the conditional distribution of its respective parameter:
\begin{equation*}
 q(\mathbf{Z}^1, \mathbf{Z}^2, \boldsymbol{\tau}, \boldsymbol{\Theta}) =  q(\boldsymbol{\Theta}| \boldsymbol{\zeta^1}, \boldsymbol{\zeta}^2)\prod^N_{i=1} q(\boldsymbol{\tau}_i | \boldsymbol{\gamma}_i) \prod^N_{j=1} q(\mathbf{Z}_{ij}^1 |  \boldsymbol{\phi}^1)q(\mathbf{Z}_{ij}^2 |  \boldsymbol{\phi}^2),
\end{equation*}
where $q(\mathbf{Z}_{ij}^1 |  \boldsymbol{\phi}^1)$and $q(\mathbf{Z}_{ij}^2 |  \boldsymbol{\phi}^2)$ are multinomial distributions, $q(\boldsymbol{\tau}_i | \boldsymbol{\gamma}_i)$ is a Dirichlet distribution,  $q(\boldsymbol{\Theta}| \boldsymbol{\zeta^1}, \boldsymbol{\zeta}^2)$ is a beta distribution,   and we have introduced the variational parameters $\boldsymbol{\phi}^1, \boldsymbol{\phi}^2, \boldsymbol{\zeta^1}, \boldsymbol{\zeta}^2$ and $\boldsymbol{\gamma}$.

%
%
Much like for an expectation-maximisation algorithm \citep{dempster77}, these parameters can be updated in a stepwise manner which iteratively optimises ${\cal L}$, and by extension $\log p(\mathbf{Y} | \boldsymbol{\beta}, \mathbf{W},  \boldsymbol{\alpha}^1, \boldsymbol{\alpha}^2)$. Updates are as follows:
\begin{eqnarray*}
\zeta^1_{gh} &=& \sum^N_{i=1} \sum^N_{j=1} \phi^1_{ijg} \phi^2_{ijh} Y_{ij} + \alpha_{gh}, \\
\zeta^2_{gh} &=& \sum^N_{i=1} \sum^N_{j=1} \phi^1_{ijg}\phi^2_{ijh} (1 -Y_{ij}) + \beta_{gh},\\
%
 \gamma_{ig} &=& \exp(\sum^P_{p=1} {\beta}_{gp} {W}_{ip}) +{\sum^N_{j=1} (\phi^1_{ijg} + \phi^2_{jig} }),\\
%
\phi^1_{ijg} & \propto & \exp\left(\Psi(\gamma_{ig}) - \Psi(\sum^G_{k=1}\gamma_{ik}) \right)\\
& \times &  \exp \left \{ \sum^G_{h=1} \phi^2_{ijh} \left [ Y_{ij}  \left( \Psi(\zeta^1_{gh}) - \Psi(\zeta^1_{gh}+\zeta^2_{gh} \right)  ) + (1-Y_{ij}) \left ( \Psi(\zeta^2_{gh}) - \Psi(\zeta^1_{gh}+\zeta^2_{gh}) \right ) \right ] \right \},\\
\phi^2_{ijg} & \propto & \exp\left(\Psi(\gamma_{jg}) - \Psi(\sum^G_{k=1}\gamma_{jk}) \right)\\
& \times &  \exp \left \{ \sum^G_{h=1} \phi^1_{ijh} \left [ Y_{ij}  \left( \Psi(\zeta^1_{hg}) - \Psi(\zeta^1_{hg}+\zeta^2_{hg} \right)  ) + (1-Y_{ij}) \left ( \Psi(\zeta^2_{hg}) - \Psi(\zeta^1_{hg}+\zeta^2_{hg}) \right ) \right ] \right \}\\
%
\end{eqnarray*}
for $i,j=1, \ldots, N$ and $g,h = 1, \ldots, G$, and where $\Psi$ denotes the digamma function \citep{abramowitz1965}.
%
%

\subsection{Estimating $\boldsymbol{\hat{\beta}}$}\label{sec:NR}
It remains to estimate $\boldsymbol{\hat{\beta}}$. Inference via a closed form solution is not possible \citep{blei03}. Instead we make use of a Newton-Raphson algorithm to maximise ${\mathcal{L}},$ by updating $\boldsymbol{\beta}^{(t+1)} =  \boldsymbol{\beta}^{(t)} -  H^{-1}\nabla$ until the algorithm has deemed to converge.
The gradient and Hessian take the following values:

\begin{eqnarray*}
\del{\beta_{iq} }{\mathcal{L}} &=& \sum^N_{n=1} W_{nq}\exp(\sum^P_{p=1} W_{np}\beta_{ip}) \\
& \times & \left \{ \Psi \left [ \sum^G_{h=1}  \exp(\sum^P_{p=1} W_{np}\beta_{hp}) \right ]  - \Psi \left [ \exp(\sum^P_{p=1} W_{np}\beta_{ip}) \right] + \Psi (\gamma_{ng}) - \Psi(\sum^G_{h=1} \gamma_{nh} )\right \}, \\
 \deltwo{\beta_{iq}}{ \beta_{jr}}{\cal L} &=& \sum^N_{n=1} W_{nq} W_{nr} \exp(\sum^P_{p=1} W_{np}(\beta_{ip} + \beta_{jp})) \left \{ \Psi^{'} \left [ \sum^G_{h=1}  \exp(\sum^P_{p=1} W_{np}\beta_{hp}) \right ]  -  {\mathbb I}_{i=j}\Psi^{'} \left [ \exp(\sum^P_{p=1} W_{np}\beta_{ip}) \right] \right \} \\
&+& {\mathbb I}_{i=j} \left ( W_{nq} W_{nr}\exp(\sum^P_{p=1} W_{np}\beta_{ip}) \right .\\
& \times & \left . \left \{ \Psi \left [ \sum^G_{h=1}  \exp(\sum^P_{p=1} W_{np}\beta_{hp}) \right ]  -  \Psi \left [ \exp(\sum^P_{p=1} W_{np}\beta_{ip}) \right] +   \Psi (\gamma_{ni}) - \Psi(\sum^G_{h=1} \gamma_{nh} )\right \} \right ).
\end{eqnarray*}

Experimental results found that the estimates obtained by the Newton-Raphson algorithm can vary wildly depending on the initial parameter settings. One strategy is to initialise the parameters using a method of moments approach proposed by \citet{minka2012} when estimating the parameters of a Dirichlet distribution. Our goal is slightly different, in that we wish to estimate the parameters $\boldsymbol{\hat{\beta}}$ with respect to the expected log of the probabilities ${\mathbb E}[\log \boldsymbol{\tau}]$ rather than the usual observed log of the probabilities. Nevertheless, the initialisation method still proves to be effective. 
In short, we initially assume that the covariates provide no additional information about the prior probability of group membership, before then setting $\beta^{(1)}_{1g} = \log \left({\mathbb E}[\delta_g] \sum^G_{h=1} \delta_h \right )$, where 
$\sum^G_{h=1} \delta_h = ({\mathbb E}[\delta_1] - {\mathbb E}[\delta_1]^2 ) / ( {\mathbb E}[\delta^2_1] - {\mathbb E}[\delta_1]^2 ).$ Intuitively, we can think of this initialisation as starting from a position of skepticism; that is, we assign weights to the covariate parameters only if it increases the lower bound. The method of moments approach serve as a reasonable initial estimate which the Newton-Raphson algorithm can then improve on.

Another difficulty which was  encountered when using the estimator experimentally was that the estimated values of coefficients for covariates with only a small number of observations tended to infinity. This may have be related to an issue known as separability in logistic regression models~\citep{Albert1984}, which typically occurs for smaller datasets, whereby for certain patterns of data points maximum likelihood estimates do not exist. While methods have been suggested to remedy this problem for logistic regression models \citep{heinze02}, as we have noted, this model is not as straightforward as other regression models, and it is not clear whether a similar approach will prove fruitful. With regards to our application in Section~\ref{sec:results}, this meant that including interaction terms proved difficult, and we were forced to omit one covariate, office location, entirely, since only five actors in the dataset practiced in one of the three locations. 

\subsection{Model Selection}
While model assumptions require the number of profiles $G$ to be fixed and known, in reality this is not the case. We therefore run the model over a range of values of $G^\prime = 1, \ldots, G^{\max}$, and compare the models post-hoc. The variational approximation to Equation (\ref{eq:ModelPost}) provides only a lower bound to the model posterior, making the use of criteria such as the Bayesian information criterion\citep{raftery1995} difficult to obtain. Other difficulties, such as determining the effective sample size of the data, also occur in this setting \citep{Hunter2008}. 
%


Alternatively, cross-validation methods can prove useful when performing model selection in a model based setting \citep{smyth2000,Hoff2008,airoldi2008}.  
In this instance the method takes the following steps:
\begin{enumerate}
\item Divide the network edges $\mathbf{Y}$ into $k$ folds of roughly equal size.
\item Drop a single fold and fit the MMESBM to the remaining data - it is straightforward to calculate $\boldsymbol{\Theta}, \boldsymbol{\tau}$ and $\boldsymbol{\delta}$. The values of $\mathbf{Z}^1$ and $\mathbf{Z}^2$ for missing edges can simply be ignored during the estimation procedure. 
\item Compare the fitted parameters against the out-of-sample data. Conditional on the fitted parameter estimates, the hold-out likelihood for an out of sample data point $Y_{ij}$ takes the form $$ p(Y_{ij} | \boldsymbol{\hat{\Theta}}, \boldsymbol{\hat{\tau}}_i, \boldsymbol{\hat{\tau}}_j) = \sum^G_{h=1} \sum^G_{g=1} \hat{\tau}_{ig} \hat{\tau}_{jh} \hat{\Theta}_{gh}^{Y_{ij}}(1-\hat{\Theta}_{gh})^{1-Y_{ij}}. $$
\item Repeat for each fold in turn. Once this has been completed, the model with highest average hold out log-likelihood, taking into account the uncertainty in the estimation, is deemed to be most suitable. In this way we can also assess goodness of fit for the model, by e.g., checking the total predicted data against the total observed data; this is described in further detail in Section~\ref{sec:results}.
\end{enumerate}

\begin{figure}[htbp]
\begin{center}
 \begin{subfigure}[b]{0.49\textwidth}
 \includegraphics[width=0.99\linewidth]{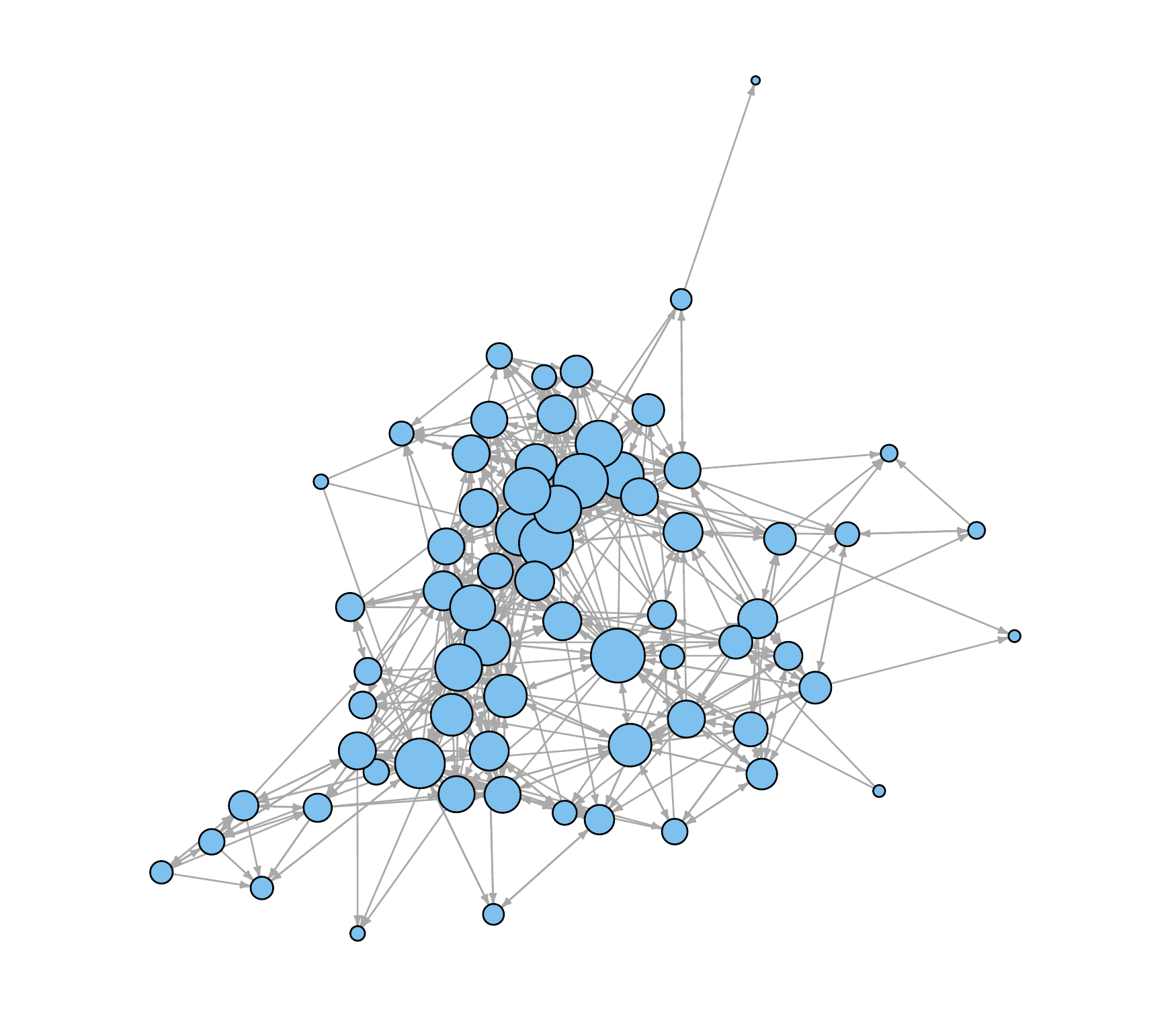}
                \vskip 0.2em
               \caption{Lazega Lawyers}
                \label{fig:Lazegaa}
         \end{subfigure}
 \begin{subfigure}[b]{0.49\textwidth}
 \includegraphics[width=0.99\linewidth]{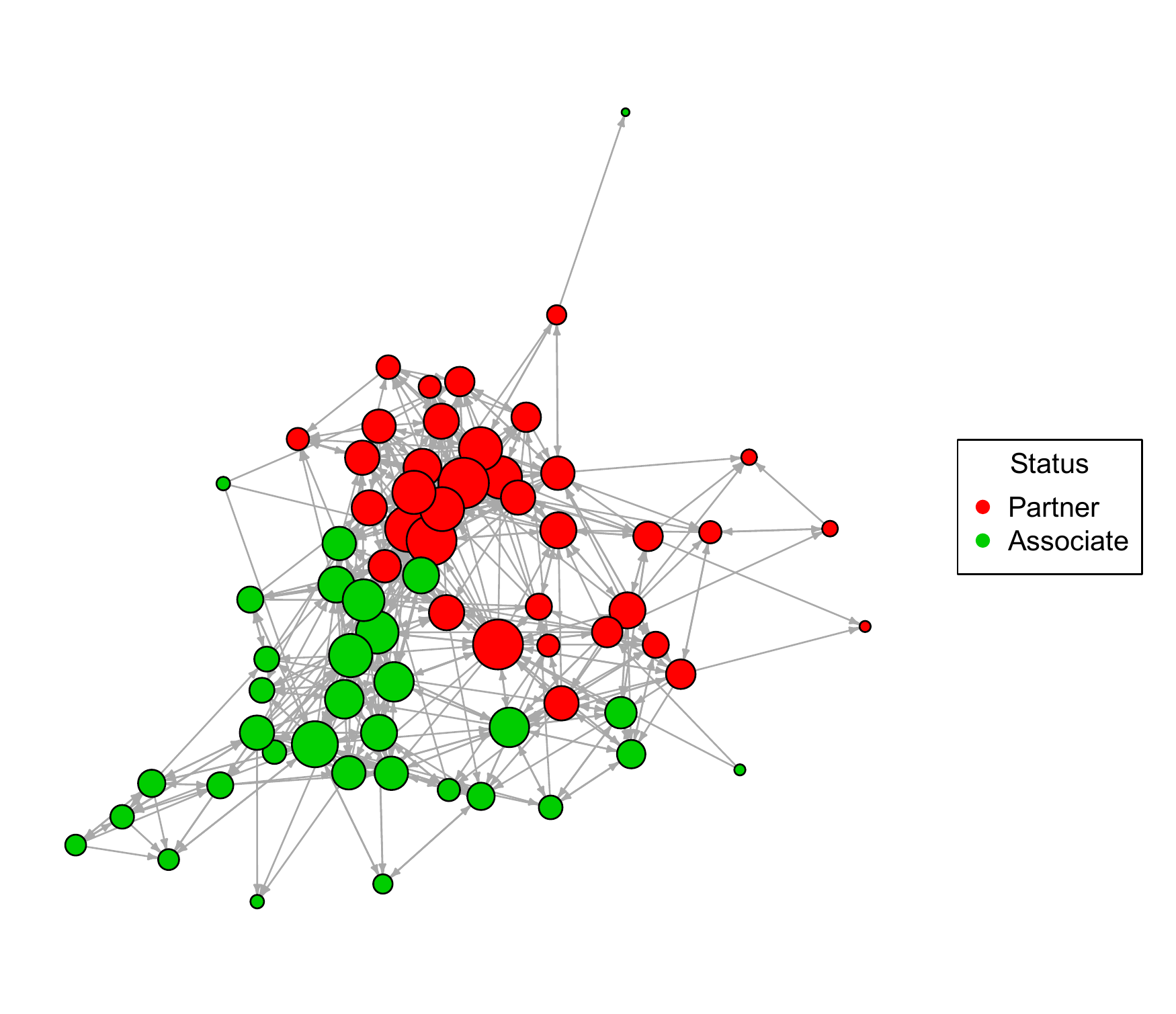}
                \vskip 0.2em
                \caption{Status}
                \label{fig:Lazegab}
         \end{subfigure}
 \begin{subfigure}[b]{0.49\textwidth}
 \includegraphics[width=0.99\linewidth]{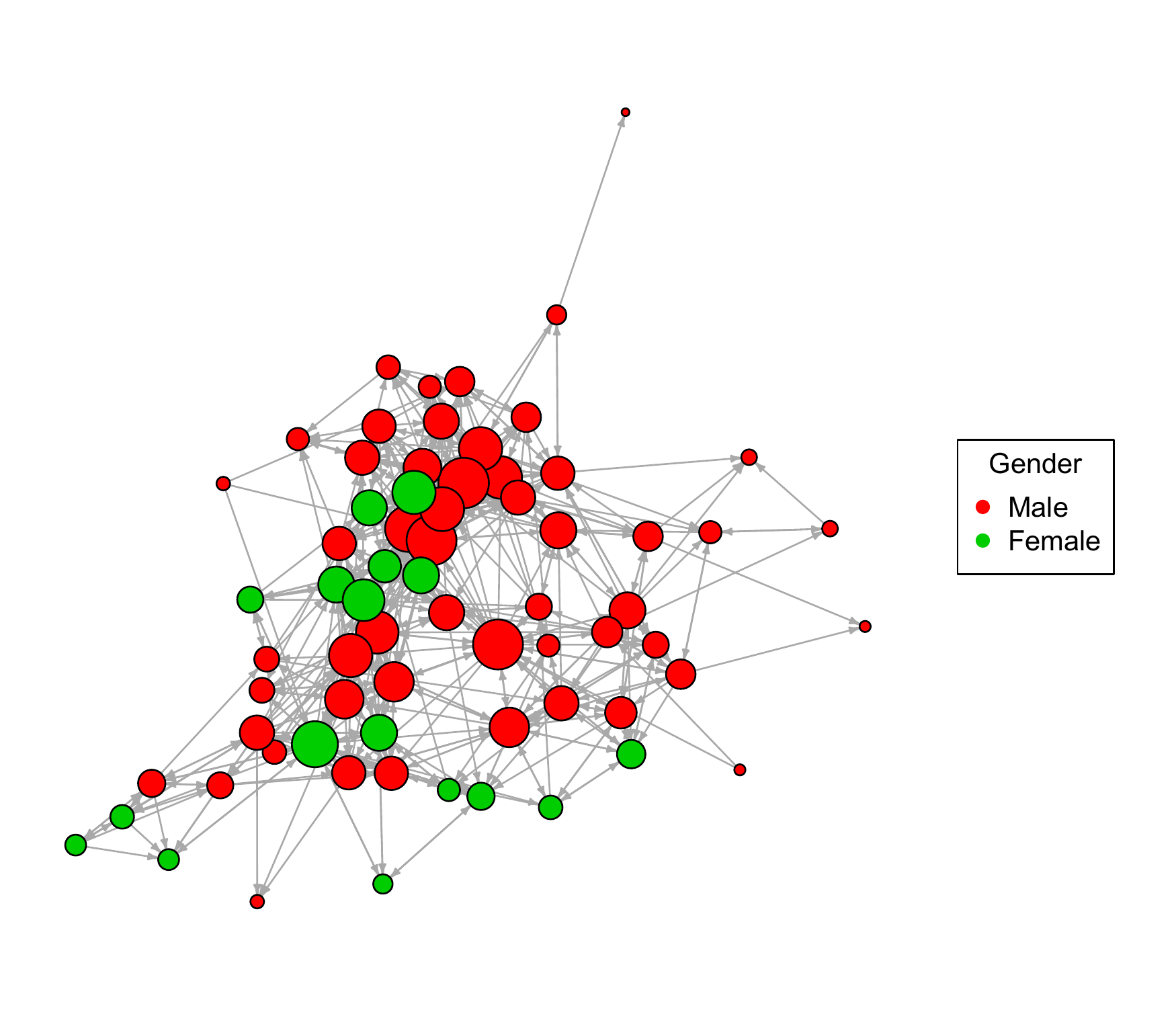}
                \vskip 0.2em
                \caption{Gender}
                \label{fig:Lazegac}
         \end{subfigure}    
\caption{Lazega Lawyers friendship network represented using a Fruchtermann-Reingold algorithm. Node size is used to give an indication of the number of links sent and received by each actor. Figures (b) and (c) colour nodes with respect to gender and status respectively. }
\end{center}
\label{fig:Lazega}
\end{figure}

\section{Lazega Lawyers Application}\label{sec:results}
We apply our method to the Lazega Lawyers dataset\footnote{The dataset is available to download at \url{http://www.stats.ox.ac.uk/ \~ snijders/siena/Lazega_lawyers_data.htm}}, obtained from a network study of corporate law partnership carried out in a Northeastern US law firm. Several features make the data of interest, the most notable being that  the lack of a strong formal working structure, coupled with large incentives to behave opportunistically create an interesting environment for the formation of network structure. Three types of network link are available from the study: strong co-worker, basic advice and friendship networks. In this paper we focus on the friendship network. Actor attribute information is also available, and described in Table \ref{tab:attribute}. These attributes have previously been incorporated into studies conducted by \citet{Gormley2010} and \citet{Snijders2006}; the former found that office location and years with the firm had a significant impact on a latent position cluster model when included as covariates for group membership, while \citet{Snijders2006} found evidence that seniority, practice and location affected the 36-actor network of partners beyond other structural effects in the data using an ERGM based approach. 

The network is visualised in Figure \ref{fig:Lazegaa} using a Fruchterman-Reingold algorithm. Two actors, who are not connected to any others in the network, are not plotted. (They are still included in the analysis.) The size of each node in the graph is representative of the overall number of links each actor has formed in the network. Note that several of the covariates are correlated. This is partly visualised in Figures \ref{fig:Lazegab} and \ref{fig:Lazegac}. These figures compare gender and status; from these figures it is clear that there are more men (53) than women (18) in the firm, and that women are more likely to be associates than partners (there are only 3 female partners in the dataset). Seniority is also highly (negatively) correlated with both age and years with the firm. In what follows the continuous attributes have been standardised to have mean$=0$ and standard deviation$=1$, to facilitate interpretation of the covariate parameters $\boldsymbol{\hat{\beta}}$.

\begin{table}[t]
\centering
\caption{Actor attribute information for Lazega Lawyers.}
\begin{tabular}{ll}
  \hline
Attribute & Description (where necessary) \\
  \hline
Seniority & Rank in firm, where 1 is the highest, 71 the lowest.\\
Status &  Indicates partner or associate in firm, with 0 = partner and 1 =associate. \\
Gender & 0 = man; 1 = woman. \\
Years with firm & \\
Age & In years.\\
Practice & 0 = litigation; 1 = corporate. \\
Law school & Yale or Harvard = 0; University of Connecticut = 1; or Other=2.\\
Office & Excluded from analysis. \\
\hline 
\end{tabular}
\label{tab:attribute}
\end{table}

\subsection{Fitting the Model}
\begin{figure}[htbp]
\begin{center}
 \begin{subfigure}[b]{0.3\textwidth}
 \includegraphics[width=0.95\linewidth]{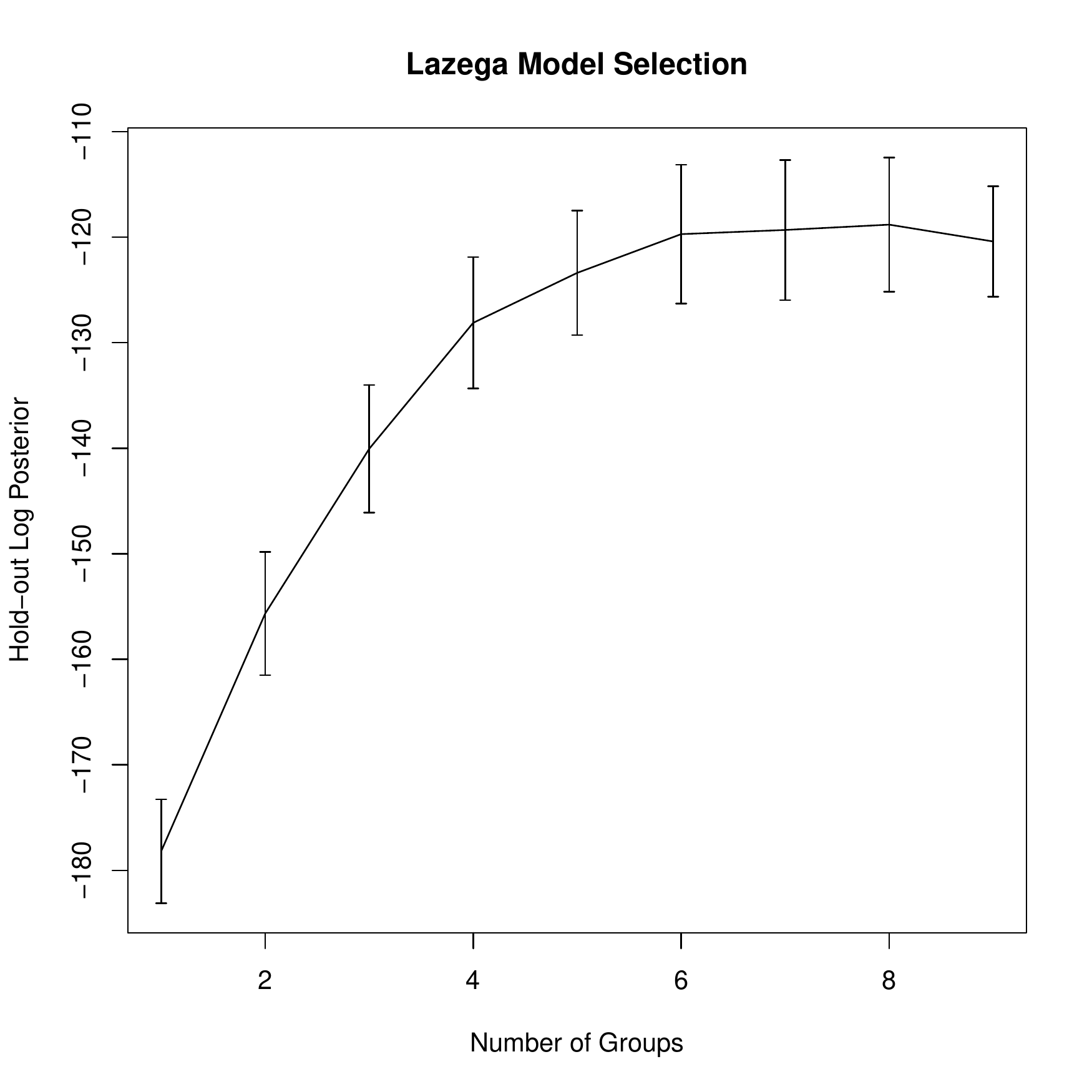}
               \caption{}
                \label{fig:ModelSelecta}
         \end{subfigure}
 \begin{subfigure}[b]{0.3\textwidth}
\centering
 \includegraphics[width=0.95\linewidth]{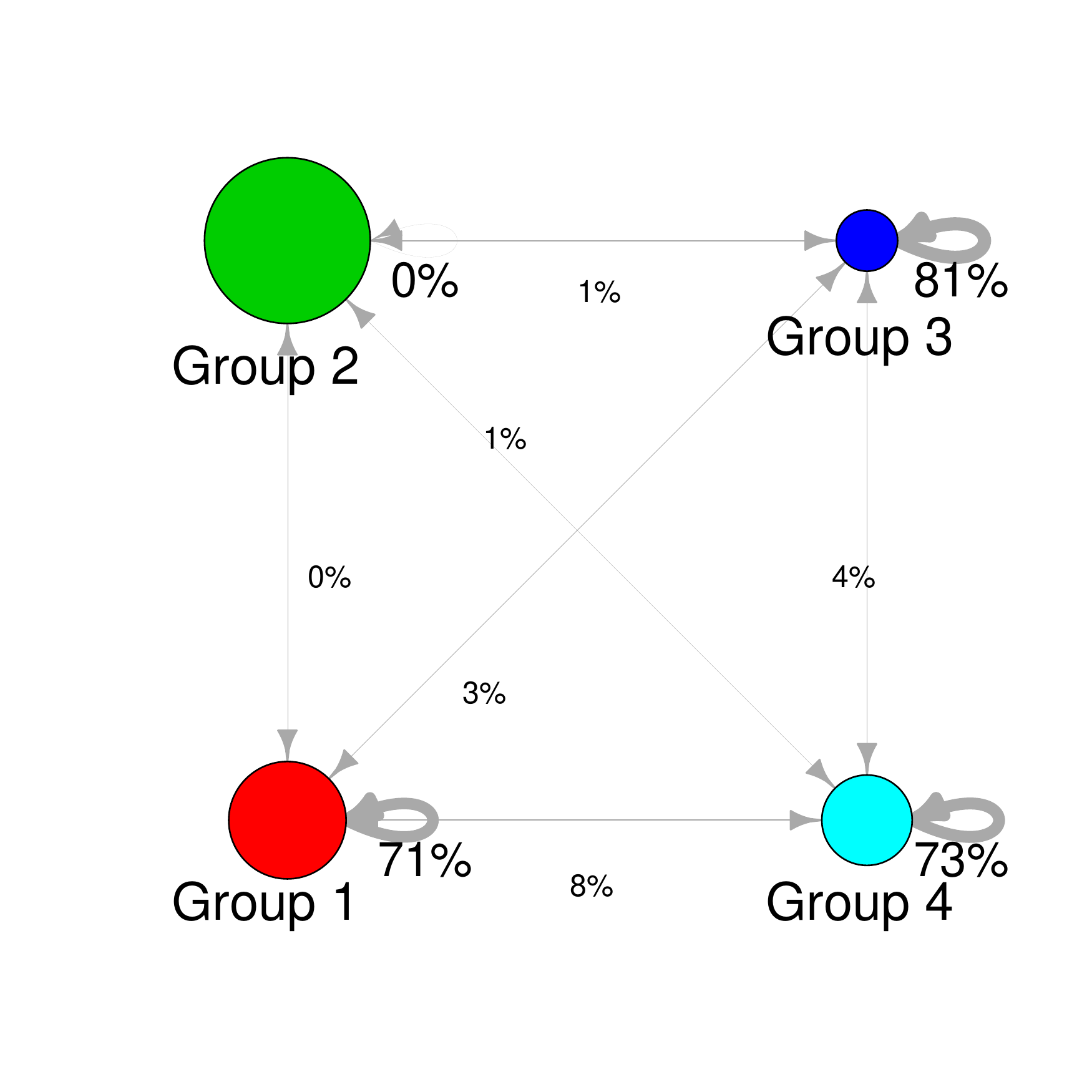}
                \caption{}
                \label{fig:ModelSelectb}
         \end{subfigure}
 \begin{subfigure}[b]{0.3\textwidth}
\centering
          \includegraphics[width=0.95\linewidth]{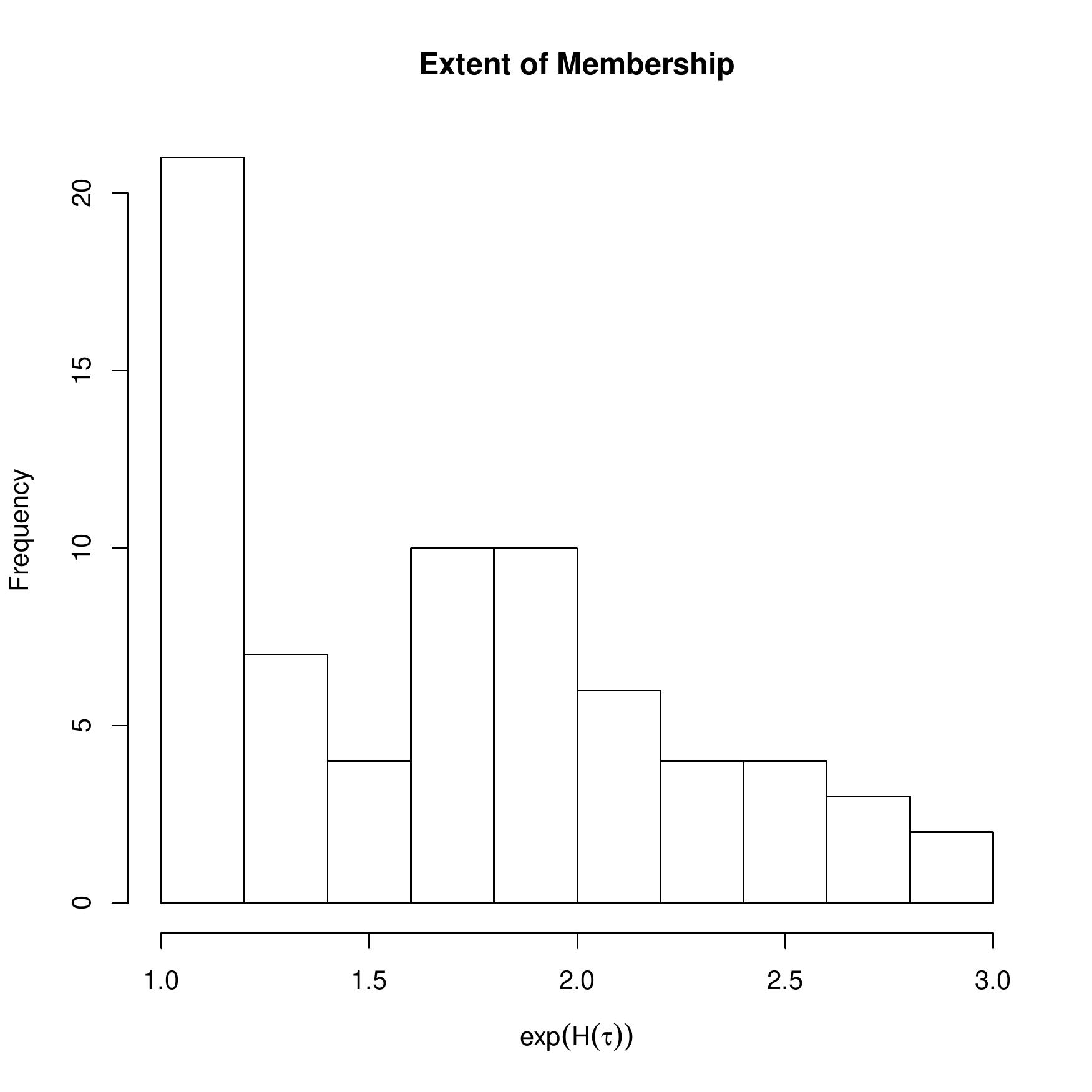}
                \caption{}
                \label{fig:ModelSelectc}
         \end{subfigure}               
\caption{Figure \ref{fig:ModelSelecta} shows the Hold-out log-likelihood for the Lazega Lawyers Friendship data. Figure \ref{fig:ModelSelectb} is a visualization of the blocks behaviour.  Figure \ref{fig:ModelSelectc} is a histogram of EoM for actors in the network. Over half the actors display mixed membership between two groups. }
\end{center}
\label{fig:ModelSelect}
\end{figure}

MMESBM models were fitted to the Lazega Lawyers data over a range of values, from $G = 1, \ldots, 9$. 
While the 10-fold cross-validated log-likelihood shown in Figure \ref{fig:ModelSelecta} is maximised for an 8 group model, the error bars for models with 4 or more groups models all overlap, suggesting a somewhat limited improvement in performance from the inclusion of additional groups. After considering the competing models, the 4 group model was deemed a satisfactory fit to the data.    

The 4 group MMESBM is represented in Figure  \ref{fig:ModelSelectb}. Each node in the diagram represents a group and is labelled accordingly. Node size reflects the overall (weighted) membership of the group, while arrow sizes loosely correspond to interaction levels, with larger arrows indicating higher levels of interaction. Selected interaction probabilities are also included in the figure. The within group interaction terms are printed in large font size beside the relevant group, while the larger of the two interaction probabilities between each pair of groups is included in smaller font size. Each between group probability is printed roughly half way between the relevant groups. 

Inspecting the figure, Groups 1, 3 and 4 can be characterised as exhibiting community-like behaviour. In each case, the probability of within group interaction occurring is far higher than would be expected in the network under the null Erd\H{o}s-R\'{e}nyi model, whereby links between actors occur independently with probability 11.5\% in this case. Group 2 is a highly antisocial group, with no interaction probabilities exceeding 1\%. The fitted values for $\boldsymbol{\hat{\Theta}}$ are provided in Table \ref{tab:thetahat}. Between group interaction occurs with low probability (less than 10\%) in all cases. 

One way to check an actor's propensity for mixed membership is to inspect their extent of profile membership (EoM) score \citep{hill73,white12}: $$\mbox{EoM}_i  = \exp (-\sum^G_{g=1}\hat{\tau}_{ig}\log\hat{\tau}_{ig}).$$ A histogram of each actor's EoM score is shown in Figure \ref{fig:ModelSelectc}. Over half (42) of the actors EoM scores are over 1.5, suggestive of at least some amount of mixed membership. Of the 20 actors with the lowest EoM score, six belong to Group 1, five to Group 3, and seven to Group 4. These actors can be viewed as being most highly involved in their respective groups and exhibit almost no mixed membership. The three actors who belong most strongly to Group 2 possess a single (received) link in the network between them. Thirteen of the fourteen actors with the highest EoM scores exhibit activity across three groups; one actor has a small amount of membership to all four groups.  With one exception these actors all have some membership of Group 2, indicating that they are not full participants in the other groups that they have membership of; only one actor appears to be highly social with Groups 1,3 and 4. 

\begin{figure}[ht!]
\begin{center}
   \begin{subfigure}[b]{0.99\textwidth}
 \includegraphics[width=0.99\linewidth]{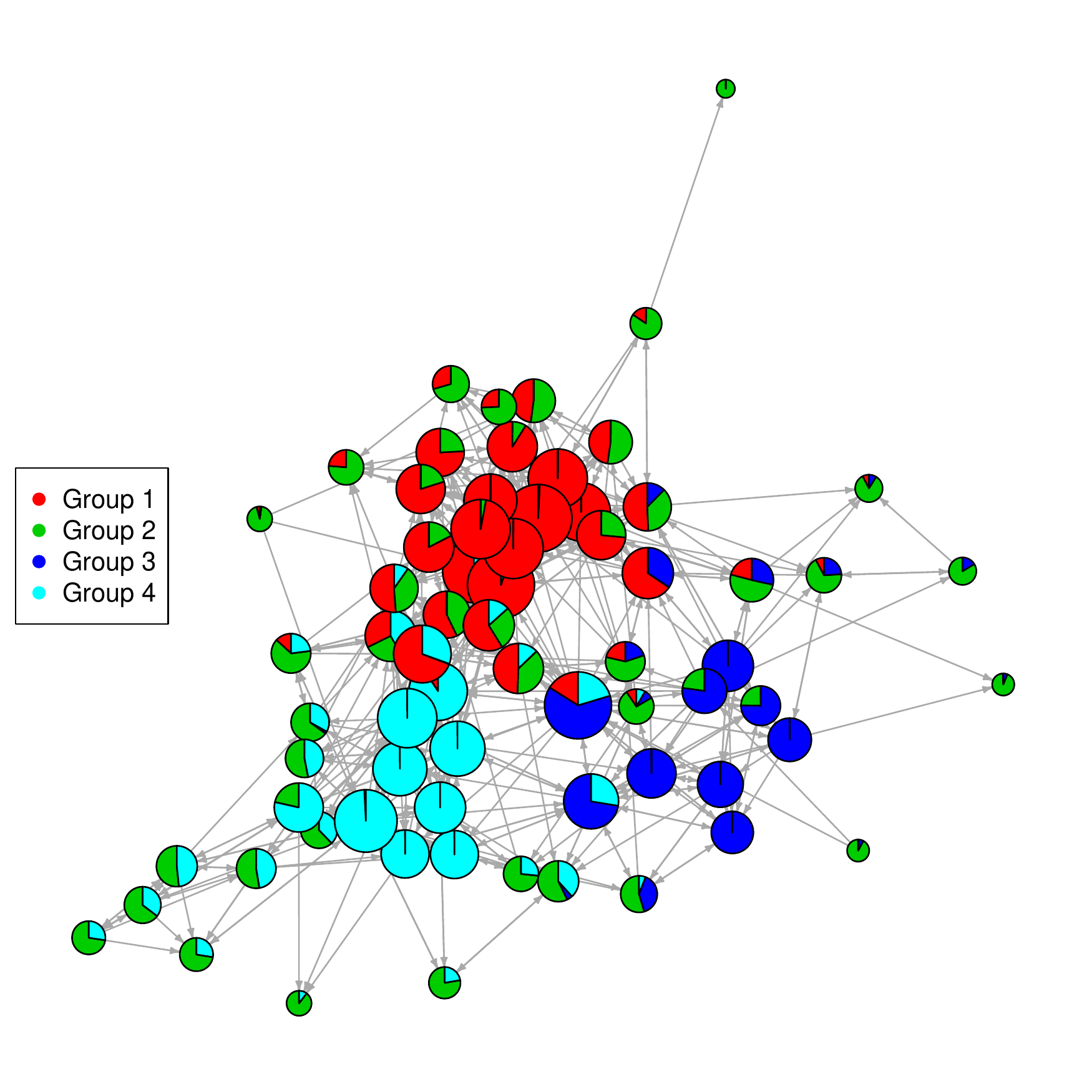}
 \caption{}
 \label{fig:Pie_Eoma}
 \end{subfigure}
\caption{Visualisation of the 4 group MMESBM fitted to the Lazega Lawyers friendship dataset.}
\end{center}
\label{fig:Pie_Eom}
\end{figure}

Figure \ref{fig:Pie_Eoma} visualises the model using the same network layout as Figures \ref{fig:Lazegaa}--\ref{fig:Lazegac}, with each of the plotted nodes assigned a pie chart representing their mixed membership to different groups. The colours in each pie chart are consistent with those in Figure~\ref{fig:ModelSelectb}. Inspecting this plot, it's clear that a large amount of mixed membership is exhibited by actors in this model, corroborating the EoM statistics reported in Figure~\ref{fig:ModelSelectc}. Recall that the size of each node in the graph is related to the popularity of the actor in question. The smaller nodes in the graph contain prominent green sections, representing Group 2,  while the largest nodes display membership to the community-like Groups 1, 3 and 4, coloured red, dark blue and light blue respectively. In particular, two types of mixed membership are occurring: actors moving between the  Groups 1, 3 and 4, and actors whose membership is split between Group 2, and another group, indicating diminished involvement. Note that the pie charts in the figure don't represent the uncertainty classification of each actor, which has been the purpose of similar plots produced by \cite{Handcock2007}. 


\begin{table}[tbh]
\centering
\caption{Estimates for blockmodel interaction $\boldsymbol{\hat{\Theta}}$.}
\begin{tabular}{lrrrrr}
 \hline
&Group 1&Group 2&Group 3&Group 4\\ 
 \hline
Group 1&0.71& 0.00& 0.01& 0.01 \\ 
Group  2&0.00& 0.00& 0.01& 0.01 \\ 
Group  3&0.03& 0.01& 0.81& 0.04 \\ 
Group  4&0.08& 0.01& 0.01& 0.73 \\ 
  \hline
\end{tabular}
\label{tab:thetahat}
\end{table}

\subsection{Covariate Parameters}
We now investigate impact of covariates in the model. Recall that the continuous attributes have been standardised to facilitate interpretation of the covariate parameters. While the Newton-Raphson step described in Section \ref{sec:NR} obtains the optimal parameter values $\boldsymbol{\hat{\beta}}$, it is necessary to obtain some estimate of the uncertainty of these parameter values before the impact of the covariates may be assessed. One approach is to consider the diagonal entries of the inverse Hessian $\mathbf{H}^{-1}$ specified in Section~\ref{sec:NR} in order to approximate the observed information matrix. The diagonal entries of this matrix should somewhat approximate the standard errors of $\boldsymbol{\hat{\beta}}$. However, this approach is limited by two facts: firstly, that we differentiate $\mathcal{L}$ and not the true log-posterior, and secondly, that we do not obtain the full Hessian matrix whereby $\mathcal{L}$ is twice differentiated with respect to all parameters, the dimension of which creates computational difficulties. Nevertheless, some information about the curvature of the parameters is obtained using this method. 

A second approach is to exploit the generative properties of the MMESBM to estimate the behaviour of $\boldsymbol{\beta}$ using a parametric bootstrapping method. Each bootstrap replication is obtained by first generating a network from the fitted model parameters using the process previously specified in Figure~\ref{fig:DataGen}. The model is then refitted to the simulated data and the results recorded. While this approach may be more reliable than the first outlined, it is worth noting that the typical under-estimation of parameter uncertainty in variational Bayes methods will be reflected in the bootstrapped values for $\boldsymbol{\beta}$. It therefore follows that while this method allows us to dismiss unimportant covariates with a high degree of certainty, care must be taken when interpreting and selecting which attributes which appear to have a meaningful impact on the network. 

Estimates of $\boldsymbol{\hat{\beta}}$ were obtained from 100 parametric bootstrap replications of the fitted model. These were in broad agreement with the estimates obtained by taking the approximate Hessian matrix. The two methods mainly disagreed on the significance of terms related to Group 3, the group with the least participation, with the Hessian term finding almost all covariate terms for this group important, whereas only two terms, the intercept, and the status of actors, appear to be meaningful based on the bootstrap estimates.

\begin{figure}[htbp]
\begin{center}
 \begin{subfigure}[b]{0.32\textwidth}
 \includegraphics[width=0.99\linewidth]{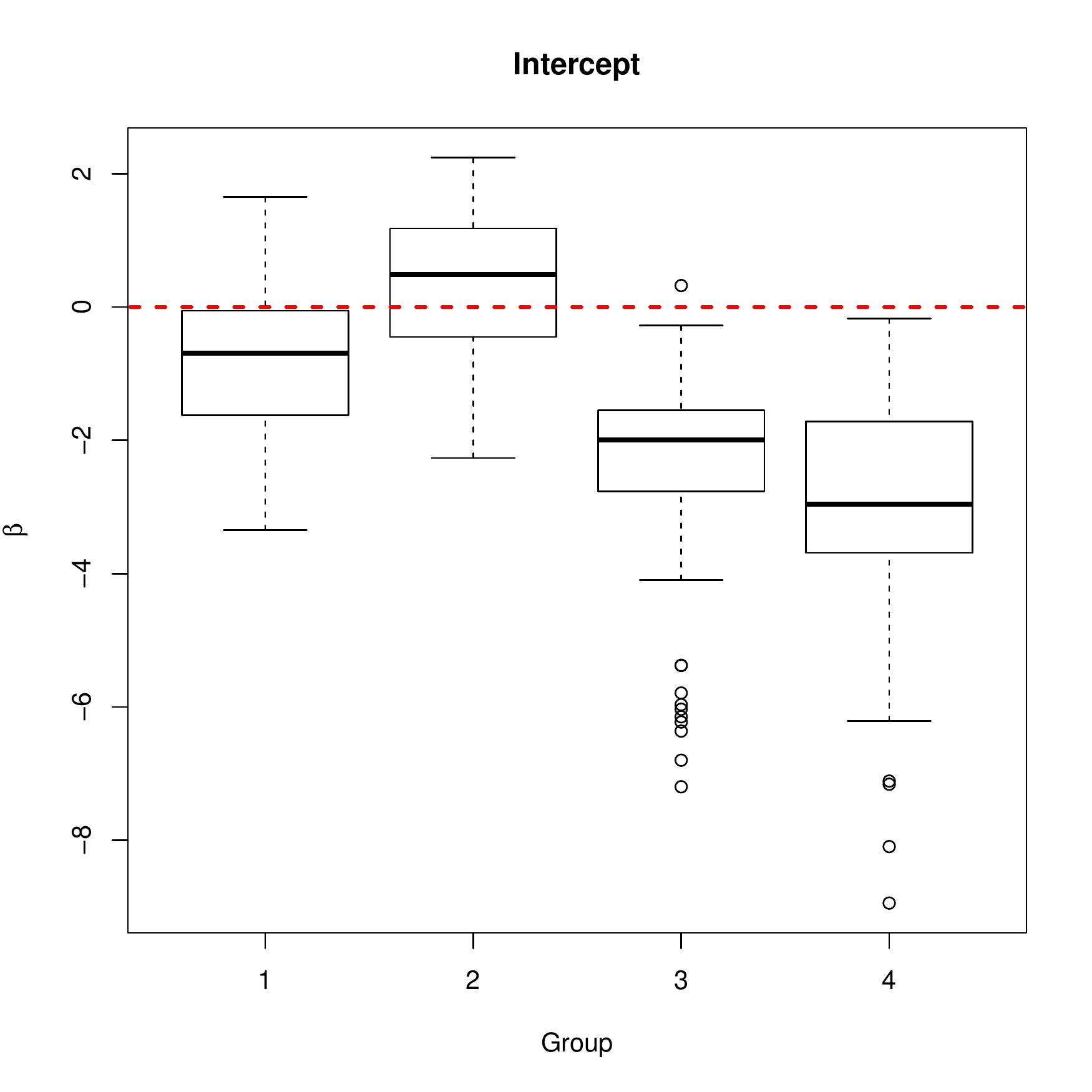}
               \caption{}
                \label{fig:Betahat_Intercept}
         \end{subfigure}
 \begin{subfigure}[b]{0.32\textwidth}
 \includegraphics[width=0.99\linewidth]{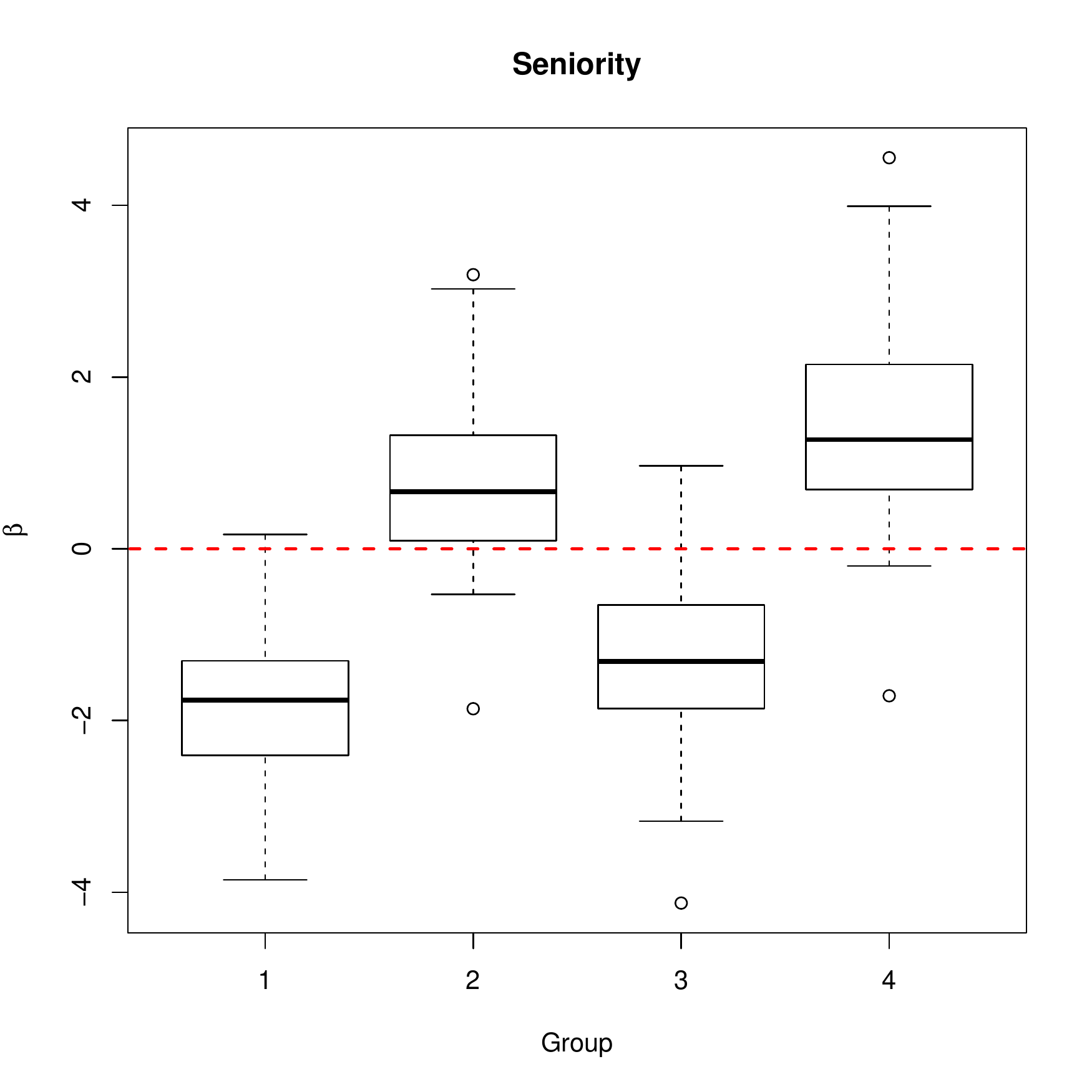}
                \caption{}
                \label{fig:Betahat_Seniority}
         \end{subfigure}
          \begin{subfigure}[b]{0.32\textwidth}
 \includegraphics[width=0.99\linewidth]{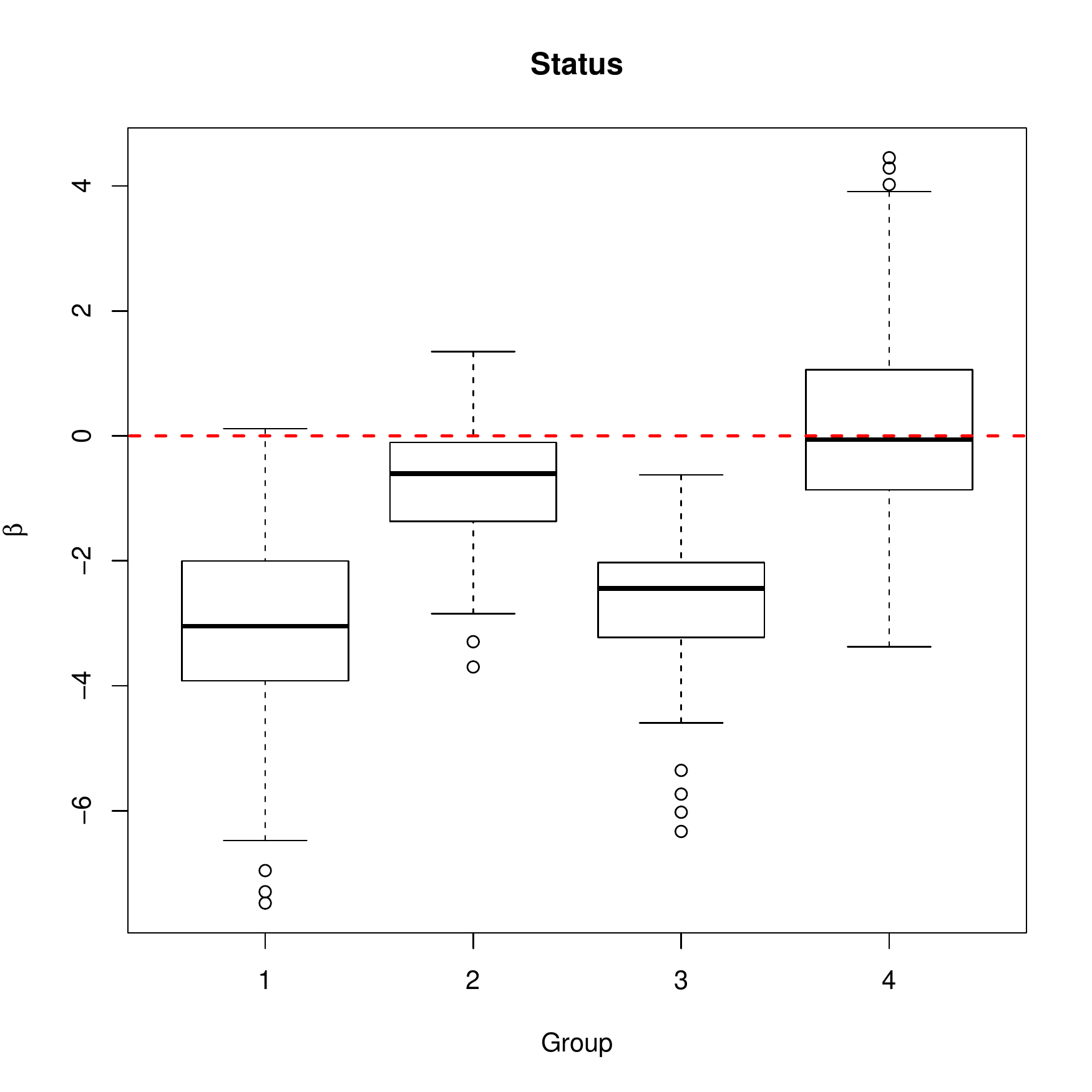}
                \caption{}
                \label{fig:Betahat_Status}
         \end{subfigure}
          \begin{subfigure}[b]{0.32\textwidth}
 \includegraphics[width=0.99\linewidth]{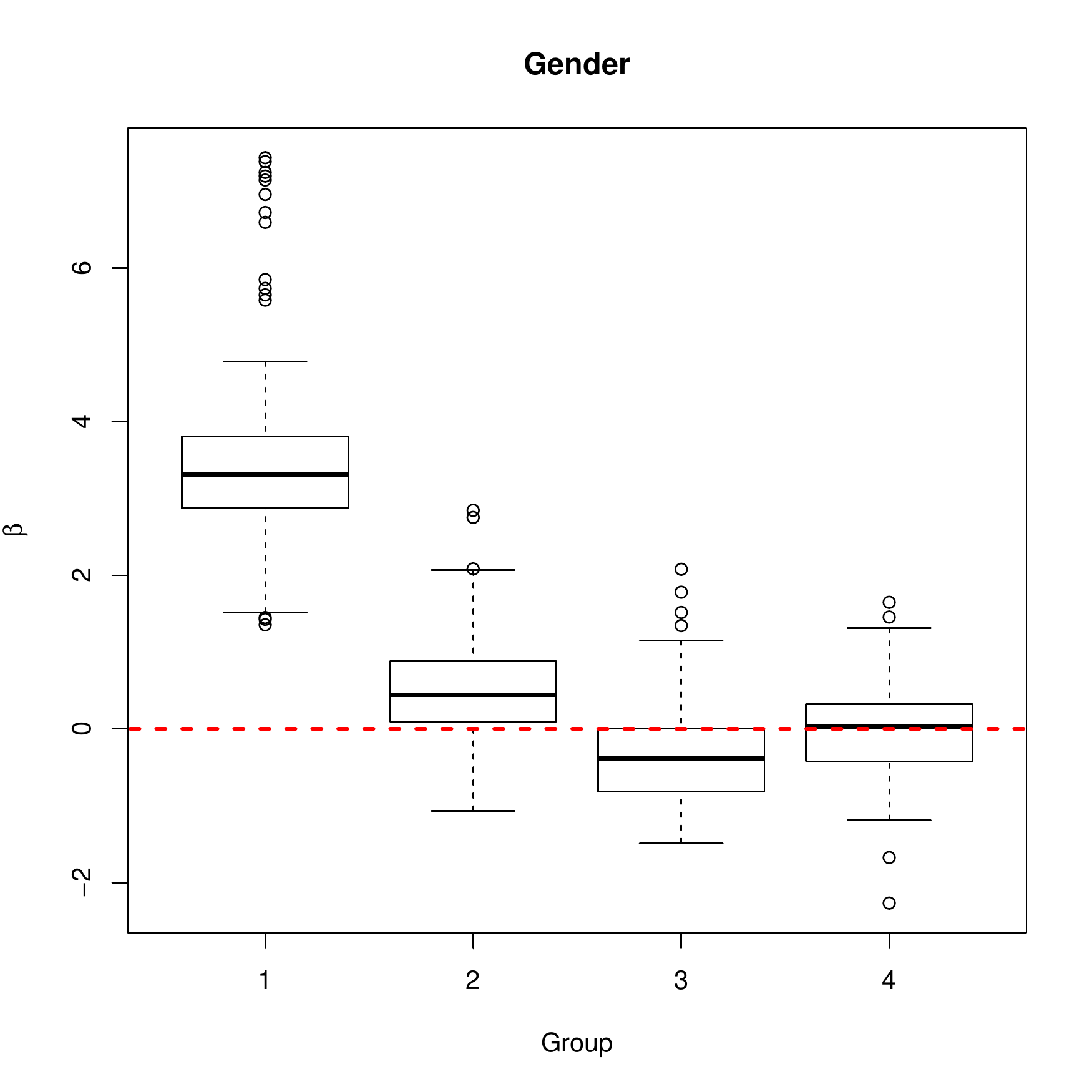}
                \caption{}
                \label{fig:Betahat_Gender}
         \end{subfigure}      
                   \begin{subfigure}[b]{0.32\textwidth}
 \includegraphics[width=0.99\linewidth]{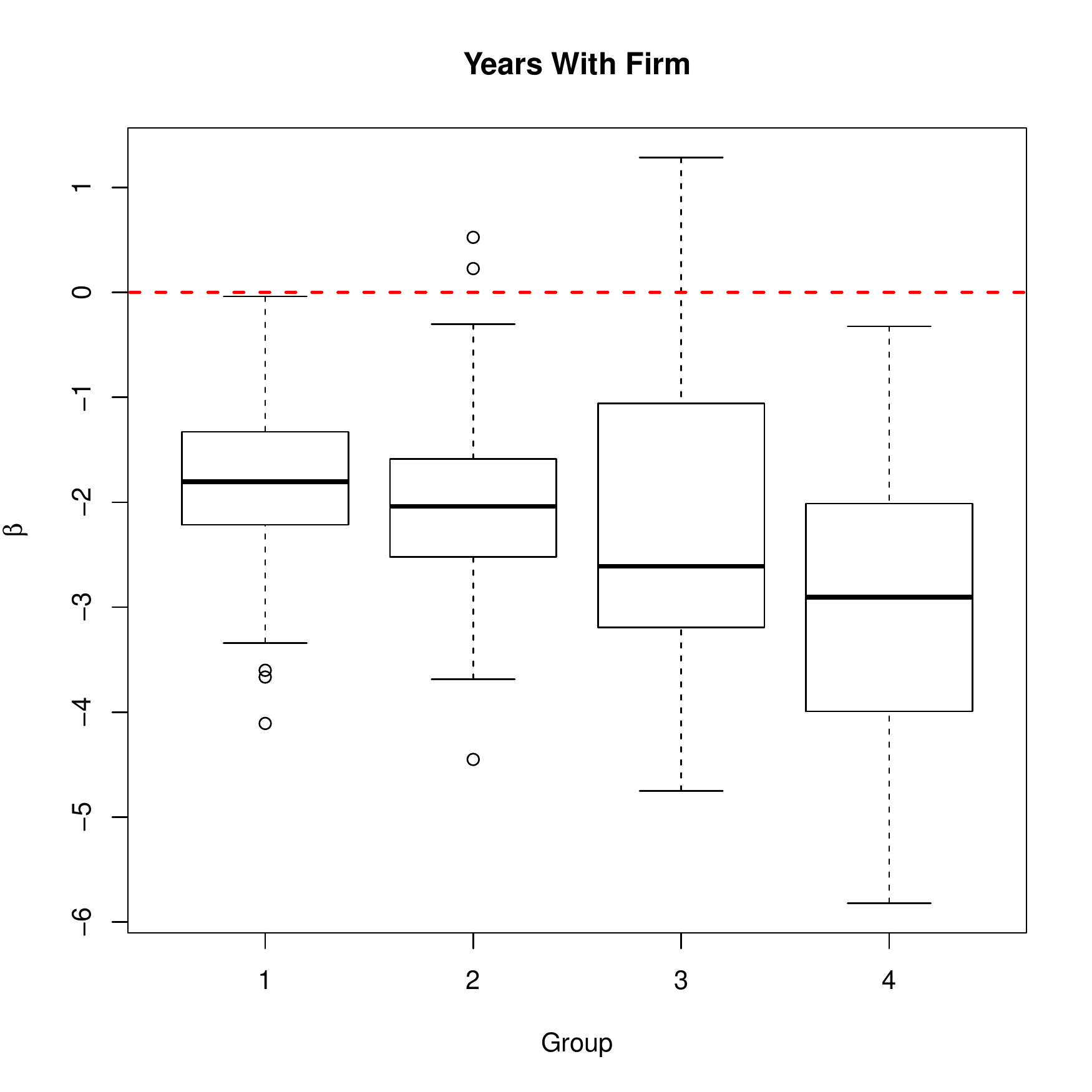}
                \caption{}
                \label{fig:Betahat_Years}
         \end{subfigure}  
                   \begin{subfigure}[b]{0.32\textwidth}
 \includegraphics[width=0.99\linewidth]{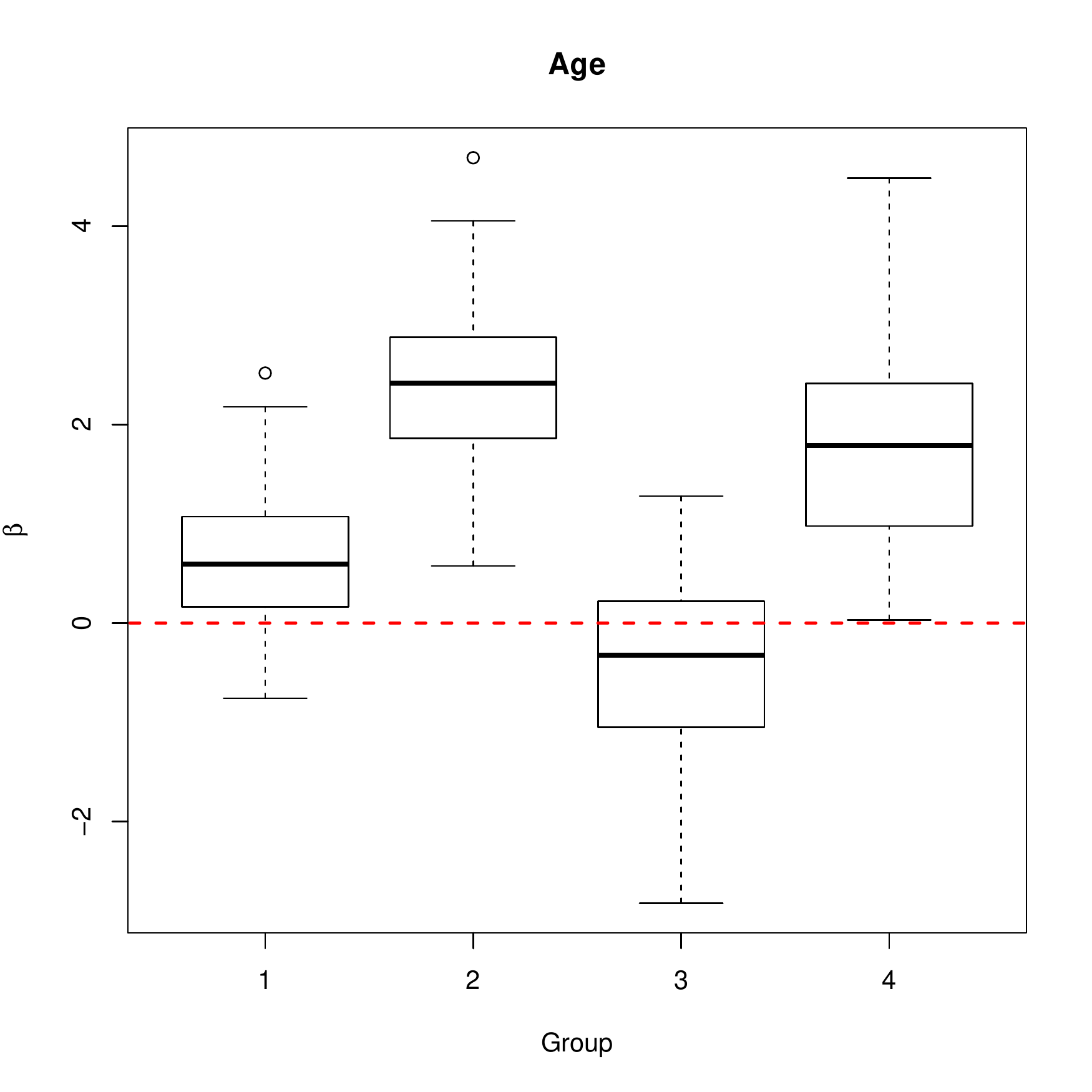}
                \caption{}
                \label{fig:Betahat_Age}
         \end{subfigure}  
                   \begin{subfigure}[b]{0.32\textwidth}
 \includegraphics[width=0.99\linewidth]{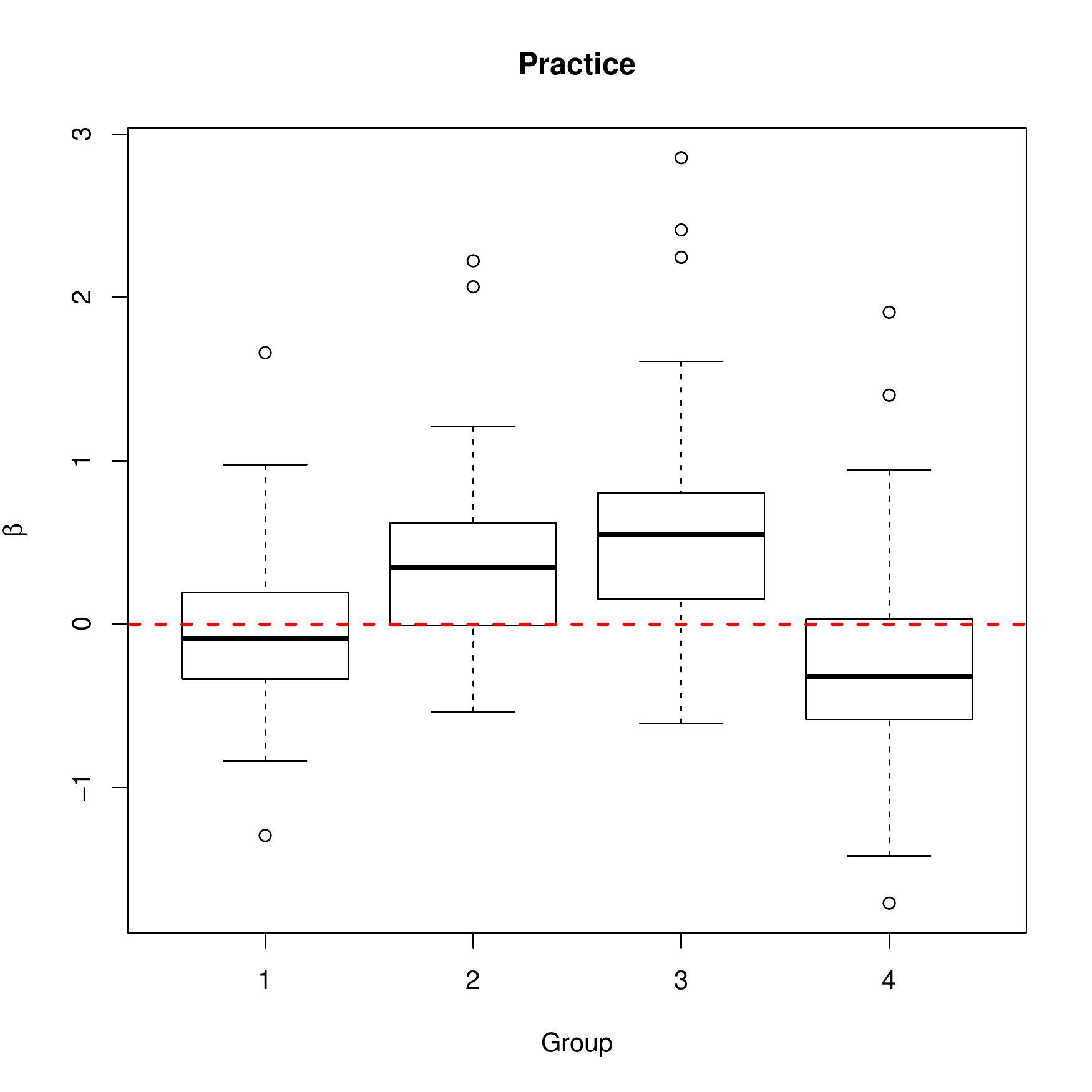}
                \caption{}
                \label{fig:Betahat_Practice}
         \end{subfigure}  
         \begin{subfigure}[b]{0.32\textwidth}
 \includegraphics[width=0.99\linewidth]{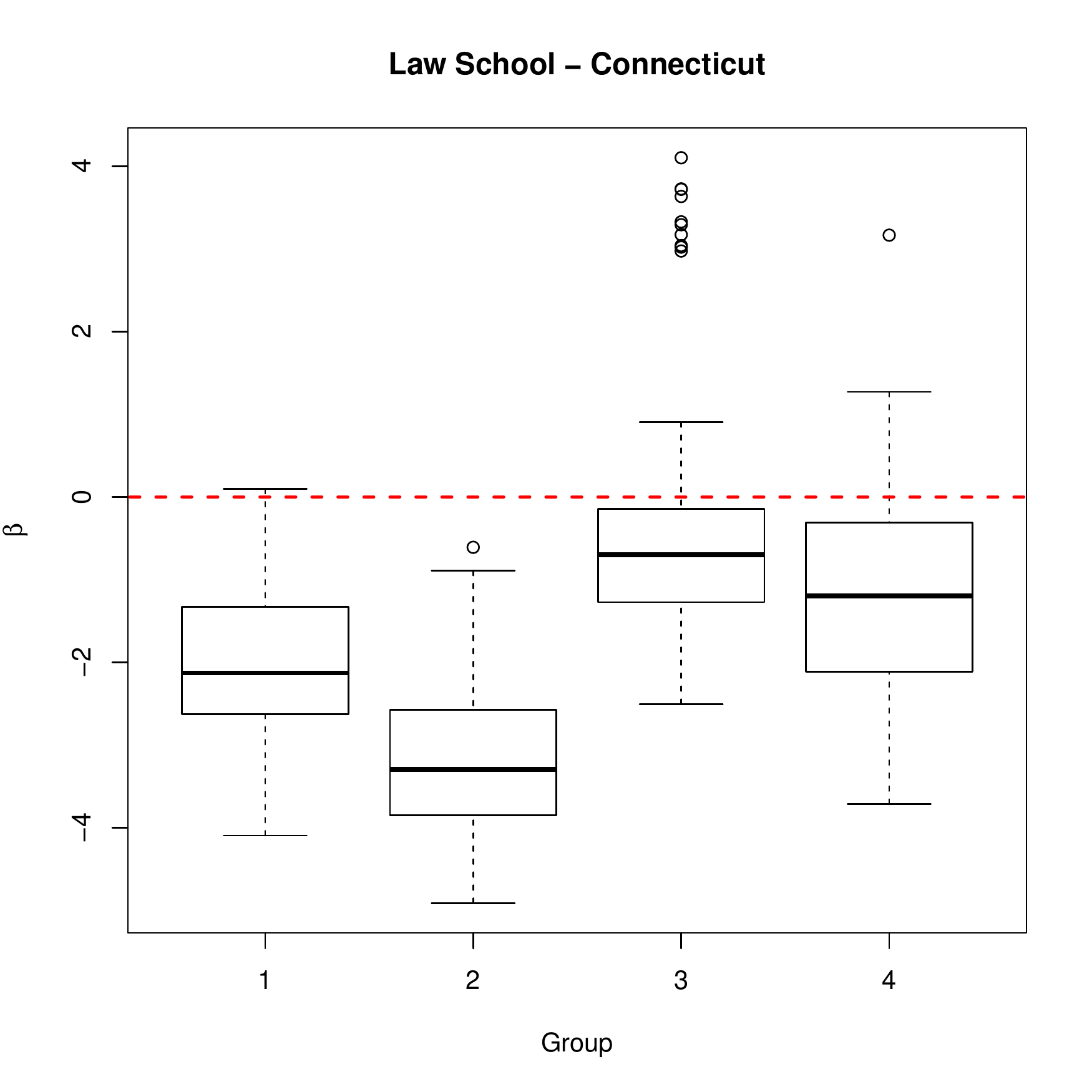}
                \caption{}
                \label{fig:Betahat_Uconn}
         \end{subfigure}  
         \begin{subfigure}[b]{0.32\textwidth}
 \includegraphics[width=0.99\linewidth]{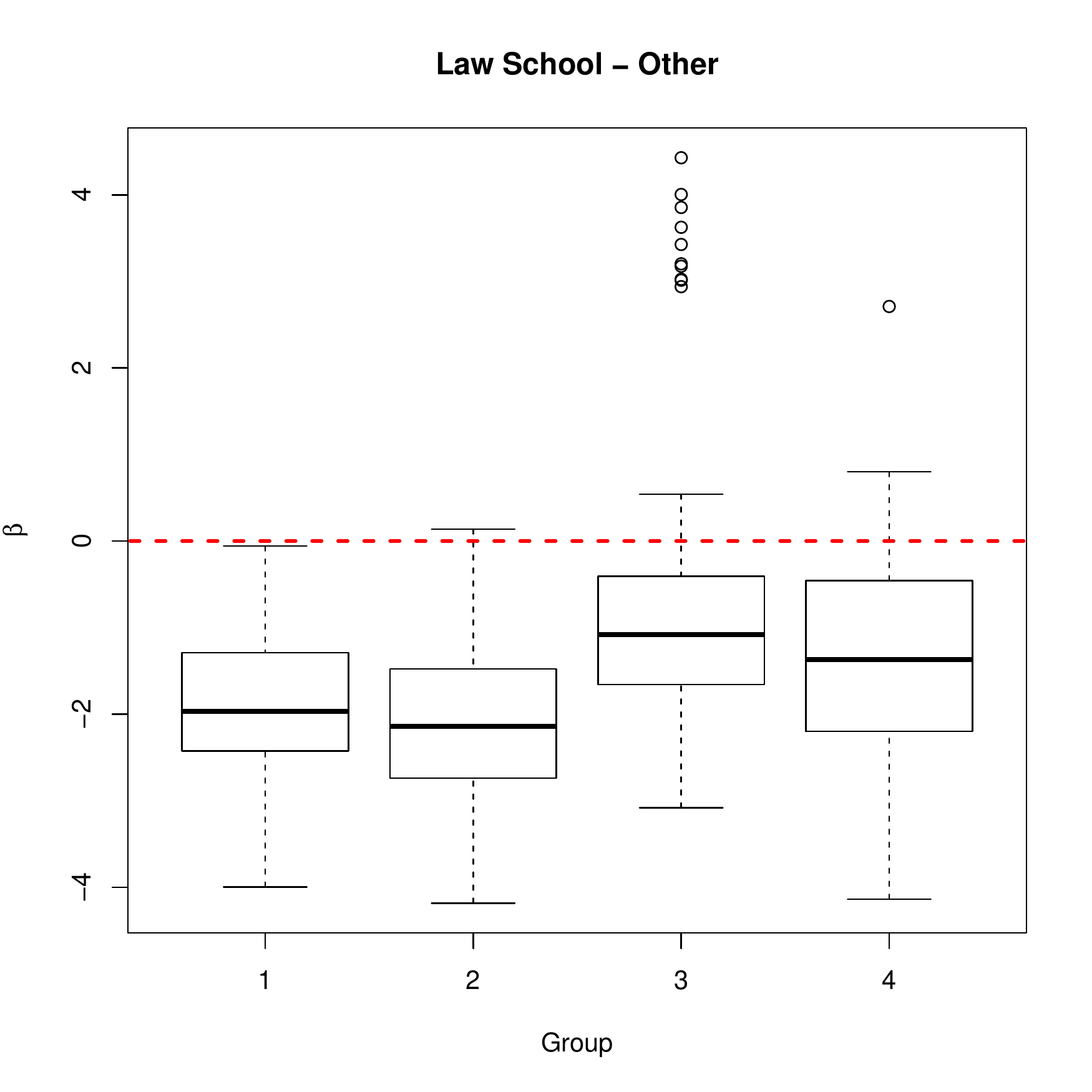}
                \caption{}
                \label{fig:Betahat_Other}
         \end{subfigure}  
\caption{Boxplots of parametric bootstrap samples of covariate parameter estimates $\boldsymbol{\hat{\beta}}$ for the 4 group MMESBM. The dashed red line occurs at zero.}
\end{center}
\label{fig:Betahat}
\end{figure}

Parameter estimates with bootstrap quantiles at the 95\% level are provided in Table \ref{tab:Betahat}, while box plots of the parametric bootstrap samples of $\boldsymbol{\hat{\beta}}$ are provided in Figures \ref{fig:Betahat_Intercept} to \ref{fig:Betahat_Other}. While our interest in the parameters directly related to covariate terms is perhaps more obvious, the behaviour of the intercepts are also worth considering; namely, intercept terms far from zero would indicate that the group membership in the network is poorly explained by the available covariate information. Of the four groups, the intercepts of Groups 3 and 4 are consistently below zero, although with quite high variance. These groups have the fewest significant covariate terms, which also suggests that their structure is only partly explained by the covariate information.

At least one covariate appears to play some part in explaining each group's structure. Gender appears to have the most sizable effect, in particular on membership to Group 1, where several other covariates are also influential, including seniority, status, years with the firm and  type of law school. It is interesting to note that despite the fact that the covariates seniority and years with the firm are negatively correlated, their respective parameters for Group 1 are in agreement. This reflects the difference in distribution between the covariates. Whereas Seniority is inherently evenly distributed across the data due to its ranked nature, the Years with Firm covariate is strongly positively skewed, reflecting the firm's tendency to recruit many junior staff and retain only the most successful. In terms of Group 1, the group consists mainly of the more senior and long established actors in the network, yet the very oldest and most 
experienced actors in the network are less involved in the network. Thus the parameter penalises the very oldest actors in the network from strong membership to Group 1, where the standardized value of Years with the Firm is further from the standardised mean (2.23 standard deviations) than Seniority (1.69 standard deviations).

A similar effect occurs in Group 2, where Years with the firm and Age disagree despite their positive correlation in the data. The difference in distribution between these covariates is less pronounced, however Age is not so strongly positively skewed as Years with the Firm. The values of the parameters mean that younger actors with relatively little experience are assigned high prior probability to Group 4, while the older actors with highest experience . In several cases, actors are assigned relatively high prior probability to both groups.

Group 3 are the group perhaps least well explained by the covariates. Almost all of the continuous covariates for actors assigned to Group 3 with high prior probability were within one standard deviation of the mean. Noticeably, however, almost all of these actors have partner status, the one covariate parameter which appears significant based on the bootstrap estimates.

While the law school parameters appear significant in this analysis, it must be noted that the upper quantiles for these parameters are close to zero, and that the variance for these parameters is large, particularly the comparison between actors attending Other law schools and those attending Harvard or Yale. If the uncertainty surrounding these terms were even slightly underestimated, then it seems likely that at least two of these terms would no longer seem significant. An exception is the negative impact which attending the law school at the University of Connecticut has on membership to Group 2 in comparison with the baseline law school of Harvard or Yale,where the impact seems to be quite large. Finally, we note that the type of practice engaged in by the actors appears to have little impact on whom they form friendships with in this model setting.

\begin{table}[t]
\centering
\caption{Estimates of the covariate parameters $\boldsymbol{\beta}$. 95\% boostrap quantile ranges (2.5\% and 97.5\%) are included in parentheses. Estimates whose quantile range does not include zero are highlighted in bold.}
\begin{tabular}{lrrrrrrrr}
  \hline
 & Group 1 & & Group 2 & & Group 3 & & Group 4 & \\ 
  \hline
Intercept & -1.62 & (-3.01,  1.00) & -0.42 & (-2.19, 1.97)  & {\bf -1.58} & (-6.30,   -0.42) & {\bf-2.42} & (-7.14,  -0.39)\\ 
  Seniority & {\bf -2.28} & (-3.61, -0.40) & 0.16 & (-0.50, 2.61) & -1.21 & (-2.82,0.49)  & 0.88 & (-0.13, 3.71)\\ 
  Status & {\bf -0.90}& (-6.73,  -0.72) & 0.02  & (-2.78, 0.88)& {\bf -1.63}& (-5.55, -1.13)& 0.91 & (-2.29,   3.97) \\ 
  Gender & {\bf 1.85 }& (1.48, 7.22)& 0.15 & (-0.55,  2.07) & -1.00 & (-1.34,   1.43)& -0.23 & (-1.14, 1.28)\\ 
  Years & {\bf -0.90} & (-3.48, -0.27) & {\bf -1.08} & (-3.57, -0.30)& -1.76 & (-4.00, 0.50)& \bf{-0.77} & (-5.28, -0.48)\\ 
  Age & -0.35 & (-0.55, 1.83) & {\bf 1.22} & (0.73 , 3.92)& -0.29 & (-2.47, 0.90)&{\bf 0.69}& (0.13, 3.68)\\ 
  Practice & 0.21& (-0.80, 0.82) & 0.38 & (-0.34, 1.20)& 0.81 & (-0.36, 1.94) & -0.66 & (-1.27, 0.71)\\ 
   UConn & {\bf -1.17} & (-3.68, -0.22) &\bf{ -2.26} & (-4.54, -0.92)& -0.78 & (-2.09, 3.68)& -0.21 &(-3.23, 1.06) \\ 
   Other & {\bf -1.15}& (-3.49, -0.11) & {\bf -1.09 }& (-3.49, -0.02)& -1.07 & (-2.86, 3.75) & -0.52 & (-3.56, 0.62)\\ 
   \hline
\end{tabular}
\label{tab:Betahat}
\end{table}

\subsection{Goodness of Fit}
\begin{figure}[htb]
\begin{center}
          \begin{subfigure}[b]{0.4\textwidth}
 \includegraphics[width=0.99\linewidth]{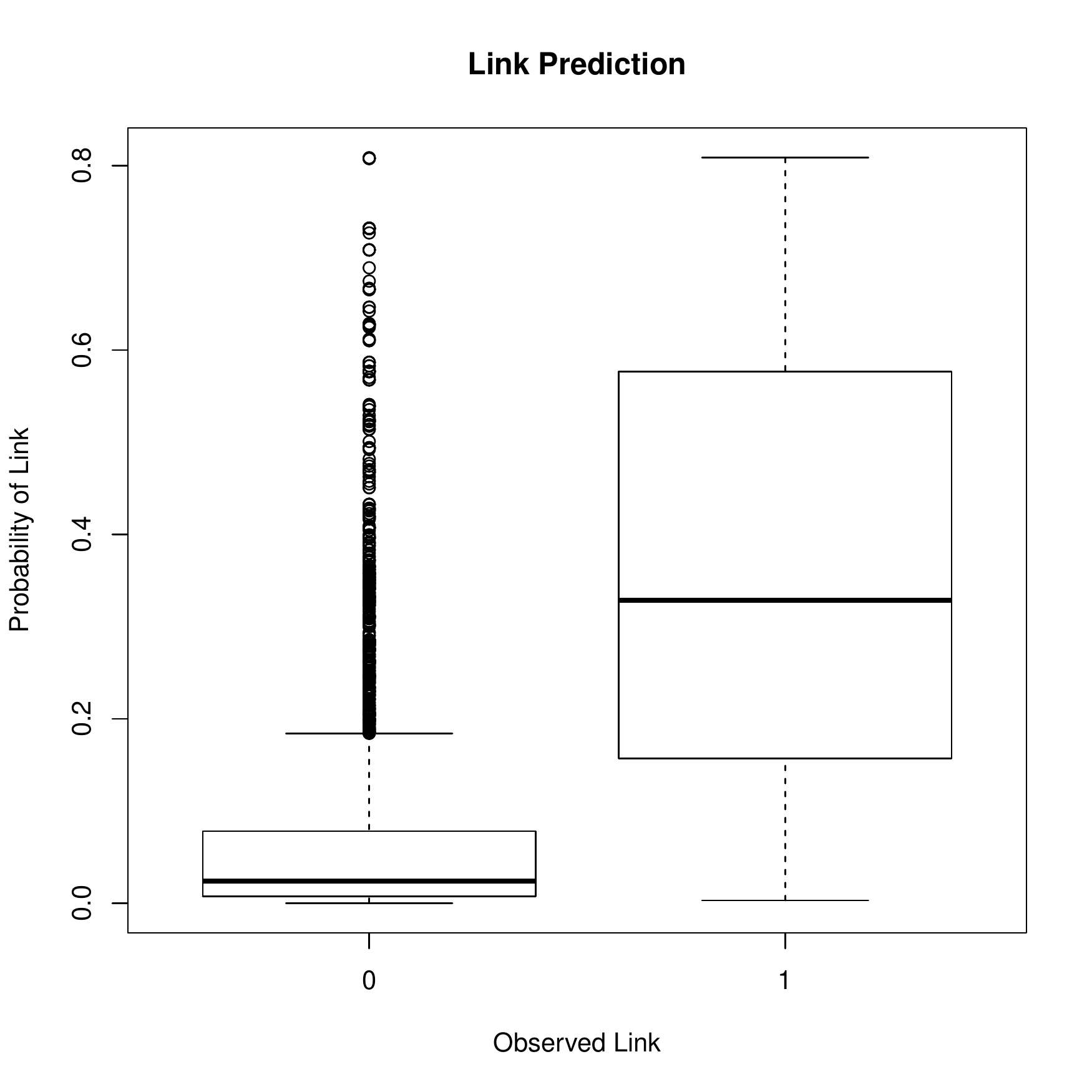}
                \caption{}
                \label{fig:obs_pred}
         \end{subfigure}     
          \begin{subfigure}[b]{0.4\textwidth}
 \includegraphics[width=1.23\linewidth]{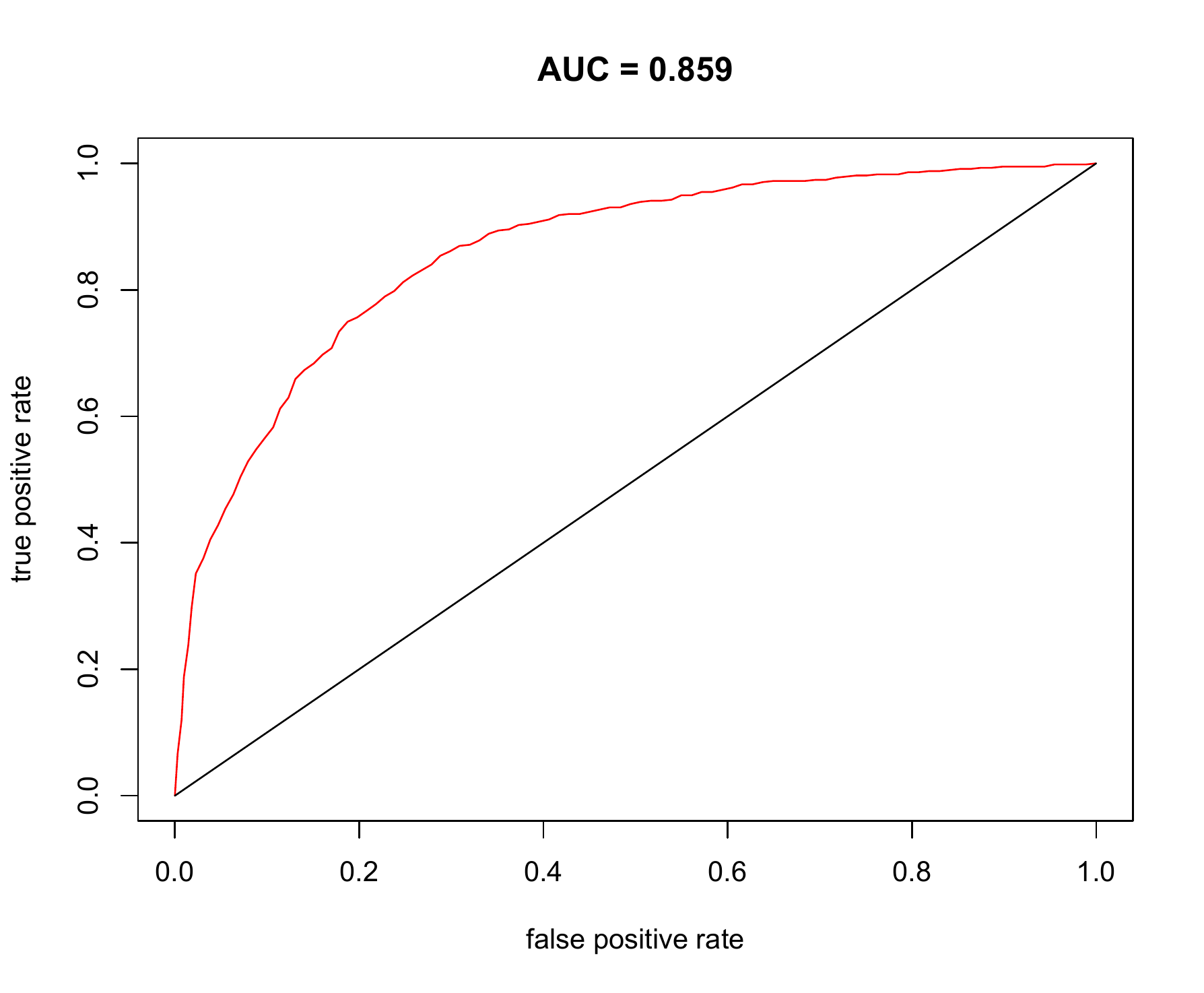}
                \caption{}
                \label{fig:AUC}
         \end{subfigure}     
\end{center}
\caption{Boxplot of link probabilities for the data based on fitted parameters. Note the high degree of separation between the probabilities for present and absent observed links. The plot on the right shows the ROC for link prediction for each of the held-out data samples during the 10-fold cross validation process. }
\label{fig:Link_Predict}
\end{figure}

Properties of the fitted model are now examined so as to determine how well the model fits the data. \citet{Hoff2002} note that one desirable property of a model is that its predictive probabilities for links and non-links be well separated. Figure~\ref{fig:obs_pred} shows boxplots of the predicted probabilities for links and non-links of the data based on the fitted parameters outlined in Section~\ref{sec:results}. The two boxplots show a high degree of separation, with the lower quartile of the observed link probabilities at roughly the same level as the top whisker of the observed non-links. Another approach is to evaluate how well the data predicts links in a hold-out modelling approach using a receiver operating characteristic (ROC) curve \citep{Hoff2008}. This is shown in Figure~\ref{fig:AUC}. Again, the model appears to perform quite well, with a total area under the curve (AUC) score of almost 0.86.

Another approach to checking model fit is based on network simulation \citep{Hunter2008,Krivitsky2008,SalterTownshend2013}. The main idea is to generate networks based on the fitted model parameters and then compare properties of these simulated networks to the observed network. Network properties which are not directly based on model parameters are considered the best indicators of model fit~\citep{Hunter2008}.  Here the model performs less well than suggested by the link prediction measures. 

\begin{figure}[htbp]
\begin{center}
 \begin{subfigure}[b]{0.31\textwidth}
 \includegraphics[width=0.99\linewidth]{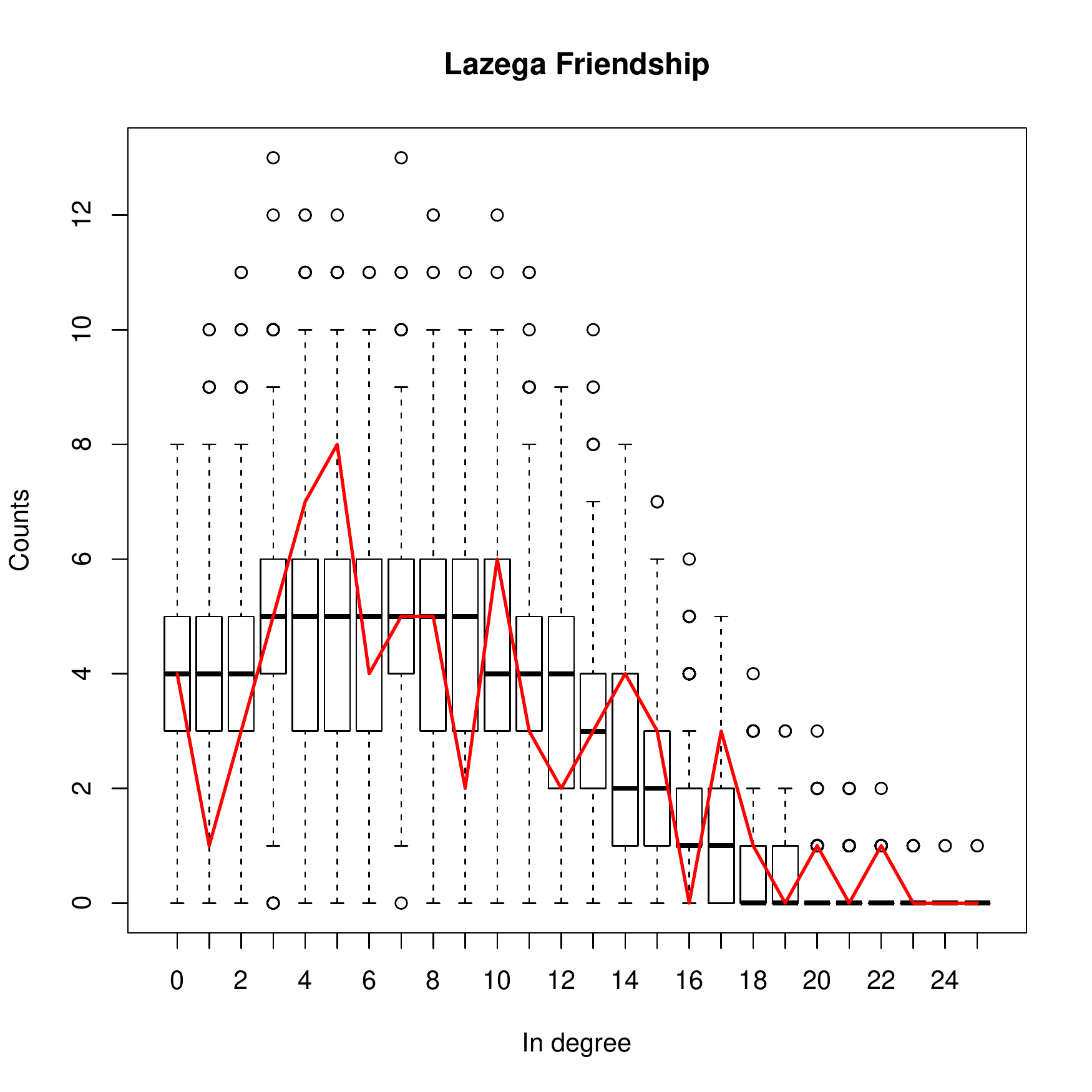}
               \caption{}
                \label{fig:InDegGOF}
         \end{subfigure}
 \begin{subfigure}[b]{0.31\textwidth}
 \includegraphics[width=0.99\linewidth]{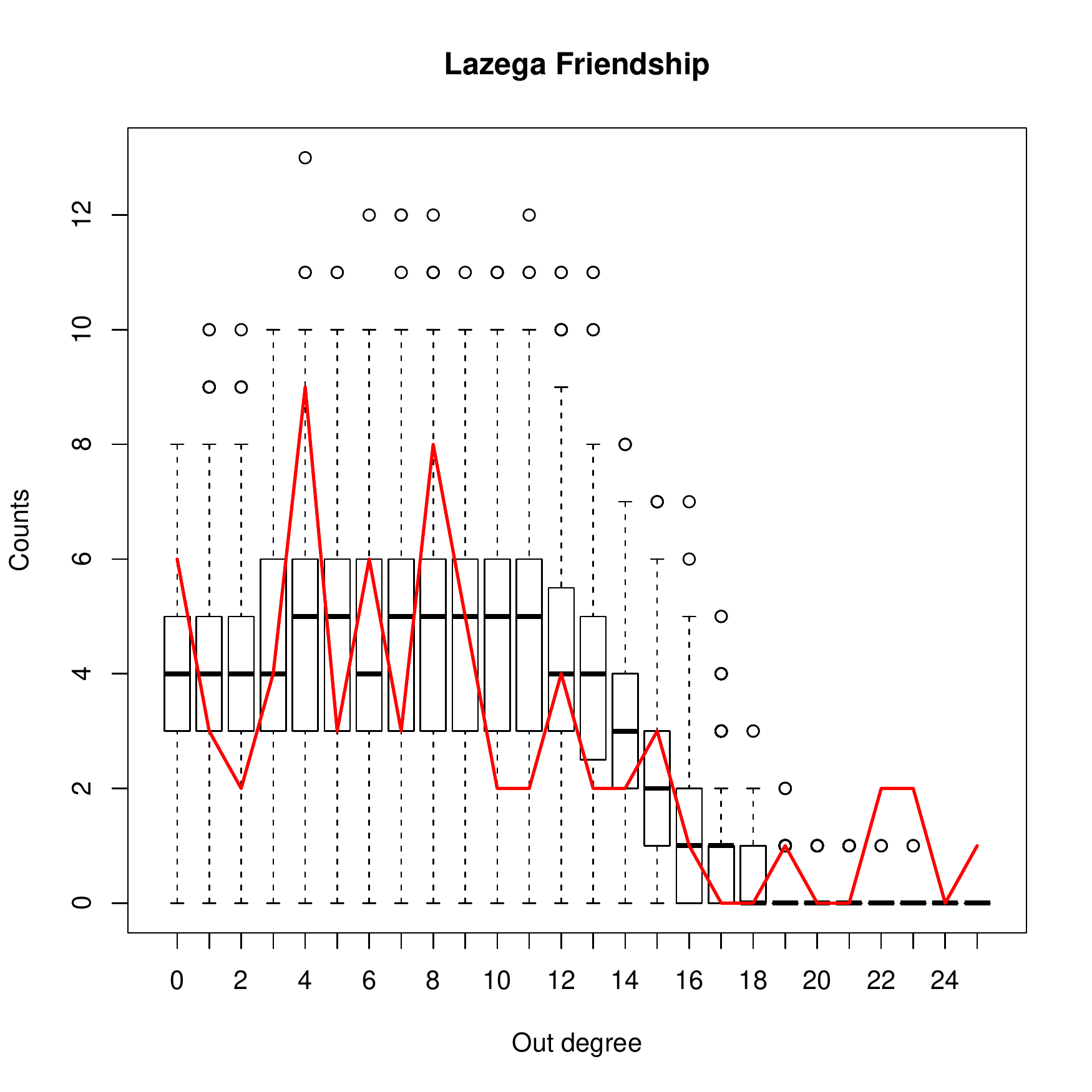}
                \caption{}
                \label{fig:OutDegGOF}
         \end{subfigure}
          \begin{subfigure}[b]{0.31\textwidth}
 \includegraphics[width=0.99\linewidth]{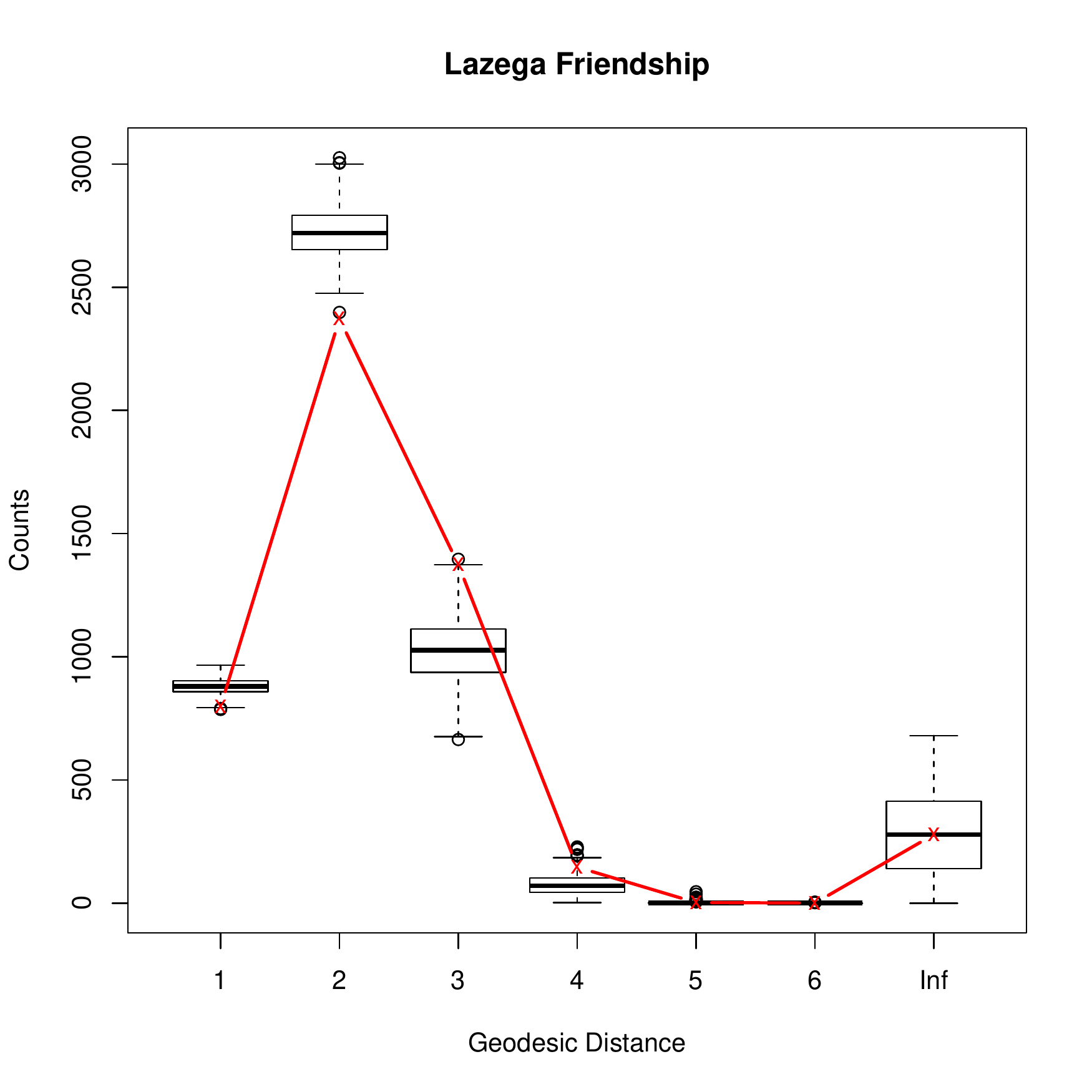}
                \caption{}
                \label{fig:GeoGOF}
         \end{subfigure}    
\caption{Goodness of fit diagnostics for the 4 group MMESBM. }
\end{center}
\label{fig:GOF}
\end{figure}

We compare the simulated networks to the observed network with respect to the following summary statistics: in degree, out degree and geodesic distance. Plots of these statistics are shown in Figures~\ref{fig:InDegGOF} to \ref{fig:GeoGOF}. These show the observed network summary statistics as a red line superimposed over boxplots of the same statistics obtained from 100 simulations. While the general behaviour of the statistics is reasonably well accounted for, the upper and lower quartiles of the in and out degree statistics appear to be too narrow, indicating a lack of variability in the simulated data. The simulated networks also fail to account for the actors in the network with the highest in and out degree. For minimum geodesic distance, while the model correctly predicts that the majority of actors are connected by two degrees of separation, it overestimates this number while underestimating the number of actors connected by three degrees of separation, again suggesting a lack of variability. This may be caused by the underestimation of uncertainty in the data generative process caused by the variational Bayes approximation already discussed in Section~\ref{sec:inference}.

\section{Conclusion}\label{sec:conclusion}
The large number of network models which have recently been introduced and extended provide the analyst with ever more tools with which to analyse relational data. In the future, it may be of interest to combine the MMESBM with other extensions to MMSBM, such as the dynamic MMSBM of \citet{Xing2010}. It is interesting to note the flexibility of the mixed membership approach, beyond allowing actors to interact in multiple social circles; for example, a majority of actors in the Lazega Lawyers dataset were assigned partial membership to Group 2, a group characterised by low interaction. This can interpreted as the model accounting for degree heterogeneity in the dataset, which must be explicitly modelled for \citep{Krivitsky2009} when using a latent position model.

In this paper an approach to incorporate actor covariates into the MMSBM has been introduced and demonstrated on a dataset. While the variational Bayes method is an effective method for inference, at least from a computational and clustering perspective, in its directly implemented form its computational cost is still of order $O(N^2)$. As currently implemented, the algorithm took several minutes to fit a single model to the Lazega Lawyers data. Model selection methods such as the outlined cross-validation approach provided a further computational burden. The case-control approximated likelihood approach introduced by \citet{raftery2011} for latent space models, which has been successfully applied in a variational Bayes setting by \citet{SalterTownshend2013} could prove effective when fitting the model to larger networks.

Model choice remains a challenge for network and mixed membership models, within the model based clustering literature and beyond. While the hold out likelihood approach which has been used in this paper gives some idea of which group choices are most suitable for the data, a high level of uncertainty still surrounds the identification of an optimal model. Similarly, care must be taken when determining which covariates appear to impact on the data.

The MMESBM as specified here can be seen to treat the covariate parameters as nuisance parameters, when they are of as much or greater interest as the other parameters in the model. While the introduction of a hyper prior would allow for inference to be performed in a more principled manner, it would also make it much more complicated, as the conjugacy between distributions would be lost. Similarly, certain properties of the Dirichlet distribution may prove too restrictive when modelling the group membership of actors, especially with the introduction of covariates; the use of other distributions, such as a logistic normal distribution may prove useful \citep{Aitchison1982,blei07}. Again, this would lead to additional inferential complexity.

While not a particular goal of the paper, it remains unclear how to choose between progressively more complex classes of model such as the SBM and MMSBM, or whether or not to include covariates, when analysing a given dataset. Potentially, another class of model, such as the latent space or ERGM may be more suitable. \cite{Hoff2008} compares fundamentally different methods by assessing their link predictive properties on hold-out samples of data, and it may be  possible to extend the use of the hold-out likelihood method employed by this paper for model selection, not just for the number of groups but also for the class of model. This possibility  comes with the caveat that link prediction is expressfully the primary goal of the analyst, when other properties in the network may be viewed as equally or more important.

\bibliography{references.bib}

\begin{thebibliography}{47}
\expandafter\ifx\csname natexlab\endcsname\relax\def\natexlab#1{#1}\fi
\expandafter\ifx\csname url\endcsname\relax
  \def\url#1{\texttt{#1}}\fi
\expandafter\ifx\csname urlprefix\endcsname\relax\def\urlprefix{URL }\fi
\providecommand{\eprint}[2][]{\url{#2}}

\bibitem[{Abramowitz \& Stegun(1965)}]{abramowitz1965}
\textsc{Abramowitz, M.} \& \textsc{Stegun, I.~A.} (1965).
\newblock \textit{Handbook of Mathematical Functions}.
\newblock Dover Publications, 1st ed.

\bibitem[{Airoldi et~al.(2007)Airoldi, Blei, Fienberg, Goldberg, Xing \&
  Zheng}]{airoldi2007}
\textsc{Airoldi, E.~M.}, \textsc{Blei, D.~M.}, \textsc{Fienberg, S.~E.},
  \textsc{Goldberg, A.}, \textsc{Xing, E.~P.} \& \textsc{Zheng, A.~X.} (2007).
\newblock \textit{Statistical Network Analysis: Models, Issues and New
  Directions}, vol. 4503 of \textit{Lecture Notes in Computer Science}.
\newblock Berlin: Springer.

\bibitem[{Airoldi et~al.(2008)Airoldi, Blei, Fienberg \& Xing}]{airoldi2008}
\textsc{Airoldi, E.~M.}, \textsc{Blei, D.~M.}, \textsc{Fienberg, S.~E.} \&
  \textsc{Xing, E.~P.} (2008).
\newblock Mixed-membership stochastic blockmodels.
\newblock \textit{Journal of Machine Learning Research} 9 1981--2014.

\bibitem[{Aitchison(1982)}]{Aitchison1982}
\textsc{Aitchison, J.} (1982).
\newblock The statistical analysis of compositional data.
\newblock \textit{Journal of the Royal Statistical Society. Series B
  (Methodological)} 44 pp. 139--177.

\bibitem[{Albert \& Anderson(1984)}]{Albert1984}
\textsc{Albert, A.} \& \textsc{Anderson, J.~A.} (1984).
\newblock On the existence of maximum likelihood estimates in logistic
  regression models.
\newblock \textit{Biometrika} 71 pp. 1--10.

\bibitem[{Beal(2003)}]{beal03}
\textsc{Beal, M.} (2003).
\newblock \textit{Variational Algorithms for Approximate {B}ayesian Inference}.
\newblock Ph.D. thesis, University College London.

\bibitem[{Bishop(2006)}]{bishop2006}
\textsc{Bishop, C.~M.} (2006).
\newblock \textit{Pattern Recognition and Machine Learning}.
\newblock Springer.

\bibitem[{Blei \& Lafferty(2007)}]{blei07}
\textsc{Blei, D.~M.} \& \textsc{Lafferty, J.~D.} (2007).
\newblock A correlated topic model of science.
\newblock \textit{Annals of Applied Statistics} 1 17--35.

\bibitem[{Blei et~al.(2003)Blei, Ng \& Jordan}]{blei03}
\textsc{Blei, D.~M.}, \textsc{Ng, A.~Y.} \& \textsc{Jordan, M.~I.} (2003).
\newblock Latent {D}irichlet allocation.
\newblock \textit{Journal of Machine Learning Research} 3 993--1022.

\bibitem[{Breiger(1974)}]{Breiger1974}
\textsc{Breiger, R.~L.} (1974).
\newblock The duality of persons and groups.
\newblock \textit{Social Forces} 53 pp. 181--190.
\newblock \urlprefix\url{http://www.jstor.org/stable/2576011}.

\bibitem[{Daudin et~al.(2008)Daudin, Picard \& Robin}]{Daudin2008}
\textsc{Daudin, J.-J.}, \textsc{Picard, F.} \& \textsc{Robin, S.} (2008).
\newblock A mixture model for random graphs.
\newblock \textit{Statistics and Computing} 18 173--183.
\newblock 10.1007/s11222-007-9046-7,
  \urlprefix\url{http://dx.doi.org/10.1007/s11222-007-9046-7}.

\bibitem[{Dempster et~al.(1977)Dempster, Laird \& Rubin}]{dempster77}
\textsc{Dempster, A.~P.}, \textsc{Laird, N.~M.} \& \textsc{Rubin, D.~B.}
  (1977).
\newblock {Maximum Likelihood from Incomplete Data via the {EM} Algorithm}.
\newblock \textit{Journal of the Royal Statistical Society B} 39 1--38.

\bibitem[{Erd\H{o}s \& R\'{e}nyi(1959)}]{Erdos1959}
\textsc{Erd\H{o}s, P.} \& \textsc{R\'{e}nyi, A.} (1959).
\newblock On random graphs {I}.
\newblock \textit{Publicationes Mathematicae Debrecen} 6 290--297.

\bibitem[{Erosheva et~al.(2007)Erosheva, Fienberg \& Joutard}]{erosheva07}
\textsc{Erosheva, E.~A.}, \textsc{Fienberg, S.~E.} \& \textsc{Joutard, C.}
  (2007).
\newblock Describing disability through individual-level mixture models for
  multivariate binary data.
\newblock \textit{The Annals of Applied Statistics} 1 502--537.

\bibitem[{{Fellows} \& {Handcock}(2012)}]{fellows2012}
\textsc{{Fellows}, I.} \& \textsc{{Handcock}, M.~S.} (2012).
\newblock {Exponential-family Random Network Models}.
\newblock \textit{ArXiv e-prints} .

\bibitem[{Gormley \& Murphy(2010)}]{Gormley2010}
\textsc{Gormley, I.~C.} \& \textsc{Murphy, T.~B.} (2010).
\newblock A mixture of experts latent position cluster model for social network
  data.
\newblock \textit{Statistical Methodology} 7 385 -- 405.
\newblock
  \urlprefix\url{http://www.sciencedirect.com/science/article/B7CRS-4Y968K0-1/2/cc7a1eb141994e86cddd3134ed5dec05}.

\bibitem[{Handcock et~al.(2007)Handcock, Raftery \& Tantrum}]{Handcock2007}
\textsc{Handcock, M.~S.}, \textsc{Raftery, A.~E.} \& \textsc{Tantrum, J.~M.}
  (2007).
\newblock Model-based clustering for social networks.
\newblock \textit{Journal of the Royal Statistical Society: Series A} 170
  1--22.

\bibitem[{Heinze \& Schemper(2002)}]{heinze02}
\textsc{Heinze, G.} \& \textsc{Schemper, M.} (2002).
\newblock A solution to the problem of separation in logistic regression.
\newblock \textit{Statistics in Medicine} 21 2409--2419.
\newblock \urlprefix\url{http://dx.doi.org/10.1002/sim.1047}.

\bibitem[{Hill(1973)}]{hill73}
\textsc{Hill, M.~O.} (1973).
\newblock Diversity and evenness: A unifying notation and its consequences.
\newblock \textit{Ecology} 54 pp. 427--432.

\bibitem[{Hoff(2008)}]{Hoff2008}
\textsc{Hoff, P.} (2008).
\newblock Modeling homophily and stochastic equivalence in symmetric relational
  data.
\newblock In J.~C. Platt, D.~Koller, Y.~Singer \& S.~Roweis, eds.,
  \textit{Advances in Neural Information Processing Systems 20}. Cambridge, MA:
  MIT Press, 657--664.

\bibitem[{Hoff et~al.(2002)Hoff, Raftery \& Handcock}]{Hoff2002}
\textsc{Hoff, P.}, \textsc{Raftery, A.} \& \textsc{Handcock, M.~S.} (2002).
\newblock Latent space approaches to social network analysis.
\newblock \textit{Journal of the {A}merican Statistical Association} 97
  1090--1098.

\bibitem[{Holland et~al.(1983)Holland, Laskey \& Leinhardt}]{holland1983}
\textsc{Holland, P.~W.}, \textsc{Laskey, K.~B.} \& \textsc{Leinhardt, S.}
  (1983).
\newblock Stochastic blockmodels: {F}irst steps.
\newblock \textit{Social Networks} 5 109 -- 137.

\bibitem[{Holland \& Leinhardt(1981)}]{Holland1981}
\textsc{Holland, P.~W.} \& \textsc{Leinhardt, S.} (1981).
\newblock An exponential family of probability distributions for directed
  graphs.
\newblock \textit{Journal of the American Statistical Association} 76 pp.
  33--50.
\newblock \urlprefix\url{http://www.jstor.org/stable/2287037}.

\bibitem[{Hunter et~al.(2008)Hunter, Goodreau \& Handcock}]{Hunter2008}
\textsc{Hunter, D.~R.}, \textsc{Goodreau, S.~M.} \& \textsc{Handcock, M.~S.}
  (2008).
\newblock Goodness of fit of social network models.
\newblock \textit{Journal of the American Statistical Association} 103
  248--258.
\newblock
  \urlprefix\url{http://ideas.repec.org/a/bes/jnlasa/v103y2008mmarchp248-258.html}.

\bibitem[{Jacobs et~al.(1991)Jacobs, Jordan, Nowlan \& Hinton}]{jacobs1991}
\textsc{Jacobs, R.~A.}, \textsc{Jordan, M.~I.}, \textsc{Nowlan, S.~J.} \&
  \textsc{Hinton, G.~E.} (1991).
\newblock Adaptive mixtures of local experts.
\newblock \textit{Neural computation} 3 79--87.

\bibitem[{Kass \& Raftery(1995)}]{raftery1995}
\textsc{Kass, R.~E.} \& \textsc{Raftery, A.~E.} (1995).
\newblock {Bayes Factors}.
\newblock \textit{Journal of the {A}merican Statistical Association} 90
  773--795.

\bibitem[{Krivitsky \& Handcock(2008)}]{Krivitsky2008}
\textsc{Krivitsky, P.~N.} \& \textsc{Handcock, M.~S.} (2008).
\newblock Fitting latent cluster models for networks with latentnet.
\newblock \textit{Journal of Statistical Software} 24 1--23.
\newblock \urlprefix\url{http://www.jstatsoft.org/v24/i05}.

\bibitem[{Krivitsky et~al.(2009)Krivitsky, Handcock, Raftery \&
  Hoff}]{Krivitsky2009}
\textsc{Krivitsky, P.~N.}, \textsc{Handcock, M.~S.}, \textsc{Raftery, A.~E.} \&
  \textsc{Hoff, P.~D.} (2009).
\newblock Representing degree distributions, clustering, and homophily in
  social networks with latent cluster random effects models.
\newblock \textit{Social Networks} 31 204--213.

\bibitem[{Kullback \& Leibler(1951)}]{Kullback1951}
\textsc{Kullback, S.} \& \textsc{Leibler, R.~A.} (1951).
\newblock On information and sufficiency.
\newblock \textit{The Annals of Mathematical Statistics} 22 pp. 79--86.

\bibitem[{Latouche et~al.(2011)Latouche, Birmel\'{e} \&
  Ambroise}]{latouche2011}
\textsc{Latouche, P.}, \textsc{Birmel\'{e}, E.} \& \textsc{Ambroise, C.}
  (2011).
\newblock Overlapping stochastic block models with application to the french
  political blogosphere.
\newblock \textit{Annals of Applied Statistics} 5 309--336.

\bibitem[{Mariadassou et~al.(2010)Mariadassou, Robin \&
  Vacher}]{mariadassou2010}
\textsc{Mariadassou, M.}, \textsc{Robin, S.} \& \textsc{Vacher, C.} (2010).
\newblock Uncovering latent structure in valued graphs: a variational approach.
\newblock \textit{The Annals of Applied Statistics} 4 715--742.

\bibitem[{McDaid et~al.(2012)McDaid, Murphy, Friel \& Hurley}]{mcdaid12}
\textsc{McDaid, A.~F.}, \textsc{Murphy, T.~B.}, \textsc{Friel, N.} \&
  \textsc{Hurley, N.} (2012).
\newblock Model-based clustering in networks with stochastic community finding.
\newblock In A.~Colubi, K.~Fokianos, E.~J. Kontoghiorghes \&
  G.~Gonz\'{a}les-Rodr\'{i}guez, eds., \textit{Proceedings of {COMPSTAT} 2012:
  20th {I}nternational Conference on Computational Statistics}. ISI-IASC,
  549--560.

\bibitem[{Minka(2012)}]{minka2012}
\textsc{Minka, T.} (2012).
\newblock Estimating a {D}irichlet distribution.
\newblock Online Manuscript.
\newblock
  \urlprefix\url{http://research.microsoft.com/en-us/um/people/minka/papers/dirichlet/}.

\bibitem[{Nowicki \& Snijders(2001)}]{Snijders2001}
\textsc{Nowicki, K.} \& \textsc{Snijders, T. A.~B.} (2001).
\newblock Estimation and prediction of stochastic blockstructures.
\newblock \textit{Journal of the {A}merican Statistical Association} 96 1077--
  1087.
\newblock
  \urlprefix\url{http://pubs.amstat.org/doi/abs/10.1198/016214501753208735}.

\bibitem[{Ormerod \& Wand(2010)}]{Ormerod10}
\textsc{Ormerod, J.} \& \textsc{Wand, M.} (2010).
\newblock Explaining variational approximations.
\newblock \textit{The {A}merican Statistician} 64 140--153.

\bibitem[{Raftery et~al.(2012)Raftery, Niu, Hoff \& Yeung}]{raftery2011}
\textsc{Raftery, A.~E.}, \textsc{Niu, X.}, \textsc{Hoff, P.~D.} \&
  \textsc{Yeung, K.~Y.} (2012).
\newblock Fast inference for the latent space network model using a
  case-control approximate likelihood.
\newblock \textit{Journal of Computational and Graphical Statistics} 21
  901--919.

\bibitem[{Robins et~al.(2006)Robins, Snijders, Wang \& Handcock}]{Robins2006}
\textsc{Robins, G.}, \textsc{Snijders, T. A.~B.}, \textsc{Wang, P.} \&
  \textsc{Handcock, M.~S.} (2006).
\newblock Recent developments in exponential random graph ($p^*$) models for
  social networks.
\newblock \textit{Social Networks} 29 192--215.

\bibitem[{Rogers et~al.(2005)Rogers, Girolami, Campbell \&
  Breitling}]{Rogers05}
\textsc{Rogers, S.}, \textsc{Girolami, M.}, \textsc{Campbell, C.} \&
  \textsc{Breitling, R.} (2005).
\newblock The latent process decomposition of c{DNA} microarray datasets.
\newblock \textit{{IEEE}/{ACM} Transactions on Computational Biology and
  Bioinformatics} 2 2005.

\bibitem[{Salter-Townshend \& Murphy(2013)}]{SalterTownshend2013}
\textsc{Salter-Townshend, M.} \& \textsc{Murphy, T.} (2013).
\newblock Variational {B}ayesian inference for the latent position cluster
  model for network data.
\newblock \textit{Computational Statistics and Data Analysis} 57 661 -- 671.

\bibitem[{Salter-Townshend et~al.(2012)Salter-Townshend, White, Gollini \&
  Murphy}]{salter12}
\textsc{Salter-Townshend, M.}, \textsc{White, A.}, \textsc{Gollini, I.} \&
  \textsc{Murphy, T.~B.} (2012).
\newblock Review of statistical network analysis: Models, algorithms, and
  software.
\newblock \textit{Statistical Analysis and Data Mining} 5 243--264.

\bibitem[{Smyth(2000)}]{smyth2000}
\textsc{Smyth, P.} (2000).
\newblock Model selection for probabilistic clustering using cross-validated
  likelihood.
\newblock \textit{Statistics and Computing} 10 63--72.

\bibitem[{Snijders(2002)}]{Snijders2002}
\textsc{Snijders, T. A.~B.} (2002).
\newblock Markov chain {M}onte {C}arlo estimation of exponential random graph
  models.
\newblock \textit{Journal of Social Structure} 3 1--40.

\bibitem[{Snijders \& Nowicki(1997)}]{Snijders1997}
\textsc{Snijders, T. A.~B.} \& \textsc{Nowicki, K.} (1997).
\newblock Estimation and prediction for stochastic bockmodels for graphs with
  latent block structure.
\newblock \textit{Journal of Classification} 14 pp.75--100.

\bibitem[{Snijders et~al.(2006)Snijders, Pattison, Robins \&
  Handcock}]{Snijders2006}
\textsc{Snijders, T. A.~B.}, \textsc{Pattison, P.~E.}, \textsc{Robins, G.~L.}
  \& \textsc{Handcock, M.~S.} (2006).
\newblock New specifications for exponential random graph models.
\newblock \textit{Sociological Methodology} 36 99--153.

\bibitem[{Wasserman \& Faust(1994)}]{Wasserman1994}
\textsc{Wasserman, S.} \& \textsc{Faust, K.} (1994).
\newblock \textit{Social network analysis: Methods and applications}.
\newblock Cambridge Univ Press.
\newblock
  \urlprefix\url{http://scholar.google.de/scholar.bib?q=info:1THqOqpVXCwJ:scholar.google.com/&output=citation&hl=de&as_sdt=0,5&ct=citation&cd=0}.

\bibitem[{White et~al.(2012)White, Chan, Hayes \& Murphy}]{white12}
\textsc{White, A.}, \textsc{Chan, J.}, \textsc{Hayes, C.} \& \textsc{Murphy,
  T.} (2012).
\newblock Mixed membership models for exploring user roles in online fora.
\newblock In N.~Ellison, J.~Shanahan \& Z.~Tufekci, eds., \textit{Proceedings
  of the {S}ixth {I}nternational {AAAI} {C}onference on {W}eblogs and {S}ocial
  {M}edia ({ICWSM} 2012)}. 599--602.

\bibitem[{Xing et~al.(2010)Xing, Fu \& Song}]{Xing2010}
\textsc{Xing, E.~P.}, \textsc{Fu, W.} \& \textsc{Song, L.} (2010).
\newblock A state-space mixed membership blockmodel for dynamic network
  tomography.
\newblock \textit{Annals of Applied Statistics} 4 535--566.
\newblock \urlprefix\url{http://projecteuclid.org/euclid.aoas/1280842130}.

\end{thebibliography}

\appendix
\section{Estimating Model Parameters}\label{sec:covar}
We know provide details for how the estimates given in Section~\ref{sec:inference} were derived. The general idea, for a given posterior with parameters $\Omega_1, \ldots, \Omega_J$, is to approximate a posterior $p(\Omega) = p(\Omega_1, \ldots, \Omega_J)$ with a set of distributions $q(\Omega_1), \ldots, q(\Omega_J)$ which can factorised independently, such that $$ p(\Omega_1, \ldots, \Omega_J) \approx q(\Omega_1) \times \cdots \times q(\Omega_J).$$
It can be shown \citep{bishop2006} that the optimal (i.e. the Kullbach-Liebler divergence minimising) form for $q(\Omega_j)$ can be found by setting 

\begin{equation*}
 q(\Omega_j) \propto \exp \left( \Esub{\log p(\Omega)}{i \neq j} \right).
\end{equation*}

We make the following approximation:

\begin{equation*}
 p(\mathbf{Z}^1, \mathbf{Z}^2, \boldsymbol{\tau}, \boldsymbol{\theta}) \approx q(\mathbf{Z}^1 | \boldsymbol{\phi}^1)q(\mathbf{Z}^2 | \boldsymbol{\phi}^2)q(\boldsymbol{\tau} | \boldsymbol{\gamma}) q(\boldsymbol{\theta} | \boldsymbol{\zeta^1}, \boldsymbol{\zeta}^2),
\end{equation*}

where we have introduced the variational parameters $\boldsymbol{\phi}^1, \boldsymbol{\phi}^2, \boldsymbol{\zeta^1}, \boldsymbol{\zeta}^2$ and $\boldsymbol{\gamma}$.

Keeping $\boldsymbol{\beta}$ fixed, and setting each $\delta_{ig} = \exp(\sum^P_{p=1}W_{ip}\beta_{gp})$, inference for $q(\boldsymbol{\tau} | \boldsymbol{\gamma})$ is as follows:
\begin{eqnarray*}
q(\boldsymbol{\tau}_i | \boldsymbol{\gamma}_i) &\propto & \exp\left\lbrace \Esub{\sum^N_{j=1} \log p(\mathbf{Z}^1_{ij} | \boldsymbol{\tau}_i) + \log p(\mathbf{Z}^2_{ji}| \boldsymbol{\tau}_i) + \log p(\boldsymbol{\tau}_i|\boldsymbol{\delta})}{\mathbf{Z}^{1}, \mathbf{Z}^{2}} \right \rbrace\\
&\propto& \prod^G_{g=1} \tau_{ig}^{\delta_{ig} - 1} \times \exp\left\lbrace \sum^N_{j=1} \sum^G_{h=1} \left ( \Esub{Z^1_{ijh}}{\mathbf{Z^1}} \log \tau_{ig} + \Esub{Z^2_{jih}}{\mathbf{Z^2}}\log \tau_{ig} \right ) \right \rbrace\\
&=& \prod^G_{g=1} \tau_{ig}^{\delta_{ig} - 1 +{\sum^N_{j=1} \Esub{Z^1_{ijg}}{\mathbf{Z^1}} + \Esub{Z^2_{jig}}{\mathbf{Z^2}} }},\\ 
\end{eqnarray*}
which we can recognise as a Dirichlet distribution. It is also straightforward to see that $q(\boldsymbol{\theta} | \boldsymbol{\zeta^1}, \boldsymbol{\zeta}^2)$ is a beta distribution:

\begin{eqnarray*}
q(\theta_{gh} | \zeta^1_{gh}, \zeta^2_{gh}) &\propto & \exp \left \lbrace \Esub{\sum^N_{i=1} \sum^N_{j=1} \log p(Y_{ij} | Z^1_{ij}, Z^2_{ij}, \theta_{gh}) + \log p(\theta | \alpha^1_{gh}, \alpha^2_{gh})}{\mathbf{Z}^1, \mathbf{Z}^2} \right \rbrace \\
&=& \theta_{gh}^{\zeta^1_{gh}}(1-\theta_{gh})^{\zeta^2_{gh}},
\end{eqnarray*}

where 
\begin{eqnarray*}
\zeta^1_{gh} &=& \sum^N_{i=1} \sum^N_{j=1} \Esub{Z^1_{ijg}}{\mathbf{Z}_{ij}^1}\Esub{Z^2_{ijh}}{\mathbf{Z^2}_{ij}} Y_{ij} + \alpha^1_{gh} \\
\zeta^2_{gh} &=& \sum^N_{i=1} \sum^N_{j=1} \Esub{Z^1_{ijg}}{\mathbf{Z}_{ij}^1}\Esub{Z^2_{ijh}}{\mathbf{Z^2}_{ij}} (1 -Y_{ij}) + \alpha^2_{gh}.
\end{eqnarray*}
Note that the calculation of $q(\boldsymbol{\tau}_i | \boldsymbol{\gamma}_i)$ did not require taking the expectation of the log posterior with respect to $\boldsymbol{\theta}$, and vice versa. This is because the parameters are conditionally independent of one another due to the presence of the indicator variable $\mathbf{Z}$. This is perhaps most clearly seen in the diagram in Figure~\ref{fig:GraphPlotb}.

Calculating $q(\mathbf{Z}^1_{ij} | \boldsymbol{\phi}^1_{ij})$ is a little trickier, since we must calculate $\Esub{\log \theta_{gh}}{\boldsymbol{\theta}}$ and $\Esub{\log \tau_{ig}}{\boldsymbol{\tau}}$:

\begin{eqnarray*}
q(\mathbf{Z}^1_{ij} | \boldsymbol{\phi}^1_{ij}) &\propto & \exp\left\lbrace \Esub{\log p(Y_{ij} | \mathbf{Z}^1_{ij}, \mathbf{Z}^2_{ij}, \boldsymbol{\theta}) + \log p(\mathbf{Z}_{ij} | \boldsymbol{\tau}_i)}{\boldsymbol{\tau}, \boldsymbol{\theta}, \mathbf{Z}^{2}} \right \rbrace\\
&=& \exp\left\lbrace \sum^G_{g=1} Z^1_{ijg} \left( \sum^G_{h=1} \Esub{Z^2_{ijh}}{\mathbf{Z^2}_{ij}} \left( Y_{ij}\Esub{\log \theta_{gh}}{\theta_{gh}}
+ (1-Y_{ij})\Esub{\log( 1 -  \theta_{gh})}{\theta_{gh}} \right)+ \Esub{\log{\tau_{ig}}}{\boldsymbol{\tau}_i} \right ) \right \rbrace\\
%
&=& \prod^G_{g=1} \left\lbrace \prod^G_{h=1} \left[ \exp(\Esub{\log \theta_{gh}}{\theta_{gh}})^{Y_{ij}} \exp(\Esub{\log(1 - \theta_{gh})}{\theta_{gh}})^{1 - Y_{ij}} \right ]^{\Esub{Z^2_{ijh}}{\mathbf{Z}_{ij}^2}} \times \exp \left ( \Esub{\log{\tau_{ig}}}{\boldsymbol{\tau}_i} \right )   \right \rbrace^{Z^1_{ijg}}.
\end{eqnarray*}

Similarly
\begin{equation*}
q(\mathbf{Z}^2_{ij} | \boldsymbol{\phi}^2_{ij}) \propto  \prod^G_{h=1} \left\lbrace \prod^G_{g=1} \left[ \exp(\Esub{\log \theta_{gh}}{\theta_{gh}})^{Y_{ij}} \exp(\Esub{\log(1 - \theta_{gh})}{\theta_{gh}})^{1 - Y_{ij}} \right ]^{\Esub{Z^1_{ijg}}{\mathbf{Z}_{ij}^1}} \times \exp \left ( \Esub{\log{\tau_{jh}}}{\boldsymbol{\tau}_j} \right )   \right \rbrace^{Z^2_{ijh}}.
\end{equation*}

We can recognise both $q(\mathbf{Z}^1_{ij} | \boldsymbol{\phi}^1_{ij}) $ and $q(\mathbf{Z}^2_{ij} | \boldsymbol{\phi}^2_{ij})$  to be multinomial distributions. 

Since the approximate distributions all have tractable form, we can calculate the required expectations, and give updates in fully parametric form:

\begin{eqnarray*}
\Esub{Z^1_{ijg}}{\mathbf{Z}_{ij}^1} &=& \phi^1_{ijg}\\
\Esub{Z^2_{ijg}}{\mathbf{Z}_{ij}^2} &=& \phi^2_{ijg}\\
 \Esub{\log{\tau_{ig}}}{\boldsymbol{\tau}_i} &=& \Psi(\gamma_{ig}) - \Psi\left(\sum^G_{k=1}\gamma_{ik}\right)\\
\Esub{\log \theta_{gh}}{\theta_{gh}} &=& \Psi(\zeta^1_{gh}) - \Psi(\zeta^1_{gh}+\zeta^2_{gh}) \\
\Esub{\log(1- \theta_{gh})}{\theta_{gh}} &=& \Psi(\zeta^2_{gh}) - \Psi(\zeta^1_{gh}+\zeta^2_{gh}).\\
\end{eqnarray*}

Parameter updates then become:

\begin{eqnarray*}
\zeta^1_{gh} &=& \sum^N_{i=1} \sum^N_{j=1} \phi^1_{ijg} \phi^2_{ijh} Y_{ij} + \alpha^1_{gh}, \\
\zeta^2_{gh} &=& \sum^N_{i=1} \sum^N_{j=1} \phi^1_{ijg}\phi^2_{ijh} (1 -Y_{ij}) + \alpha^2_{gh},\\
\gamma_{ig} &=& \delta_{ig} +{\sum^N_{j=1} (\phi^1_{ijg} + \phi^2_{jig} }) \\
\phi^1_{ijg} & \propto & \exp\left(\Psi(\gamma_{ig}) - \Psi(\sum^G_{k=1}\gamma_{ik}) \right),\\
& \times &  \exp \left \{ \sum^G_{h=1} \phi^2_{ijh} \left [ Y_{ij}  \left( \Psi(\zeta^1_{gh}) - \Psi(\zeta^1_{gh}+\zeta^2_{gh} \right)  ) + (1-Y_{ij}) \left ( \Psi(\zeta^2_{gh}) - \Psi(\zeta^1_{gh}+\zeta^2_{gh}) \right ) \right ] \right \},\\
\phi^2_{ijg} & \propto & \exp\left(\Psi(\gamma_{jg}) - \Psi(\sum^G_{k=1}\gamma_{jk}) \right)\\
& \times &  \exp \left \{ \sum^G_{h=1} \phi^1_{ijh} \left [ Y_{ij}  \left( \Psi(\zeta^1_{hg}) - \Psi(\zeta^1_{hg}+\zeta^2_{hg} \right)  ) + (1-Y_{ij}) \left ( \Psi(\zeta^2_{hg}) - \Psi(\zeta^1_{hg}+\zeta^2_{hg}) \right ) \right ] \right \}.\\
%
\end{eqnarray*}
\subsection{Estimating Covariate Parameters}
Recall that the log-posterior is intractable, and that we instead maximise a lower bound ${\cal L}$:
\begin{equation*}
{\cal L} = \Esub{ \log p(\mathbf{Y}, \mathbf{Z}^1, \mathbf{Z}^2, \boldsymbol{\tau}, \boldsymbol{\theta} | \boldsymbol{\alpha}^1, \boldsymbol{\alpha}^2, \boldsymbol{\delta})}{\mathbf{Z}^1, \mathbf{Z}^2, \boldsymbol{\tau}, \boldsymbol{\theta}} - \Esub{\log q(\mathbf{Z}^1, \mathbf{Z}^2, \boldsymbol{\tau}, \boldsymbol{\theta})}{\mathbf{Z}^1, \mathbf{Z}^2, \boldsymbol{\tau}, \boldsymbol{\theta}},
\end{equation*}
Where 
\begin{eqnarray*}
\Esub{ \log p(\mathbf{Y}, \mathbf{Z}^1, \mathbf{Z}^2, \boldsymbol{\tau}, \boldsymbol{\theta} | \boldsymbol{\alpha}^1, \boldsymbol{\alpha}^2, \boldsymbol{\delta})}{\mathbf{Z}^1, \mathbf{Z}^2, \boldsymbol{\tau}, \boldsymbol{\theta}} &=& \sum^N_{i=1} \sum^N_{j=1} \Esub{ \log p(Y_{ij} | \mathbf{Z}^1_{ij}, \mathbf{Z}^2_{ij}, \boldsymbol{\theta})}{\mathbf{Z}^1, \mathbf{Z}^2, \boldsymbol{\theta}}\\
& + & \sum^N_{i=1} \sum^N_{j=1}\Esub{ \log p(\mathbf{Z}^1_{ij} | \boldsymbol{\tau}_i)}{\mathbf{Z}^1, \boldsymbol{\tau}} + \Esub{ \log p(\mathbf{Z}^2_{ij} | \boldsymbol{\tau}_j)}{\mathbf{Z}^2, \boldsymbol{\tau}}\\
& + & \sum^N_{n=1}\Esub{\log p(\boldsymbol{\tau_n} | \boldsymbol{\delta)}}{\boldsymbol{\tau}} \\
&+& \Esub{\log p(\boldsymbol{\theta} | \boldsymbol{\alpha}^1,  \boldsymbol{\alpha}^2) }{\boldsymbol{\theta}},
\end{eqnarray*}
and 
\begin{eqnarray*}
\Esub{\log q(\mathbf{Z}^1, \mathbf{Z}^2, \boldsymbol{\tau}, \boldsymbol{\theta})}{\mathbf{Z}^1, \mathbf{Z}^2, \boldsymbol{\tau}, \boldsymbol{\theta}} &=& \sum^N_{i=1} \sum^N_{j=1} \Esub{\log q(\mathbf{Z}_{ij}^1 | \boldsymbol{\phi}_{ij}^1)}{\mathbf{Z}^1} + \Esub{\log q({\mathbf{Z}}_{ij}^2 | \boldsymbol{\phi}_{ij}^2)}{\mathbf{Z}^2} \\
&+& \sum^N_{n=1} \Esub{\log q(\boldsymbol{\tau_n} | \boldsymbol{\gamma_n})}{\boldsymbol{\tau}}\\
 & + &\Esub{\log q(\boldsymbol{\theta} | \boldsymbol{\zeta}^1, \boldsymbol{\zeta}^2 )}{\boldsymbol{\theta}}.
\end{eqnarray*}

This is straightforward to calculate:
\begin{eqnarray*}
\Esub{ \log p(Y_{ij} | \mathbf{Z}^1_{ij}, \mathbf{Z}^2_{ij}, \boldsymbol{\theta})}{\mathbf{Z}^1, \mathbf{Z}^2, \boldsymbol{\theta}} & = & \sum^G_{g=1} \sum^G_{h=1} \phi^1_{ijg}\phi^2_{ijh} \left \lbrace Y_{ij}(\Psi(\zeta^1_{gh}) - \Psi(\zeta^1_{gh}+\zeta^2_{gh}))  \right . \\
& + & \left . (1- Y_{ij})(\Psi(\zeta^2_{gh}) - \Psi(\zeta^1_{gh}+\zeta^2_{gh})) \right \rbrace \\
\Esub{ \log p(\mathbf{Z}^1_{ij} | \boldsymbol{\tau}_i)}{\mathbf{Z}^1, \boldsymbol{\tau}} &=& \Esub{ \sum^G_{g=1} {Z}^1_{ijg}\log{\tau}_{ig}}{\mathbf{Z}^1, \boldsymbol{\tau}} \\
&=& \sum^G_{g=1} {\phi}^1_{ijg} \times \left \lbrace \Psi({\gamma}_{ig}) - \Psi\left( \sum^G_{k=1} \gamma_{ik} \right ) \right \rbrace,\\
\Esub{ \log p(\mathbf{Z}^2_{ij} | \boldsymbol{\tau}_j)}{\mathbf{Z}^2, \boldsymbol{\tau}} &=& \sum^G_{g=1} {\phi}^2_{ijg} \times \left \lbrace \Psi({\gamma}_{jg}) - \Psi\left( \sum^G_{k=1} \gamma_{jk} \right ) \right \rbrace, \\
\Esub{\log p(\boldsymbol{\tau_n} | \boldsymbol{\delta)}}{\boldsymbol{\tau}} &=& \Esub{ \log \Gamma \left(\sum^G_{h=1} \delta_{h} \right ) - \sum^G_{k=1} \log \Gamma (\delta_{k})  + \sum^G_{g=1} (\delta_g -1)\log{\tau_{ng}}}{\boldsymbol{\tau}} \\
&=&  \log \Gamma \left(\sum^G_{h=1} \delta_{h} \right ) - \sum^G_{k=1} \log \Gamma (\delta_{k})  + \sum^G_{g=1} (\delta_g -1) \times \left \lbrace \Psi({\gamma}_{ng}) - \Psi\left( \sum^G_{k=1} \gamma_{nk} \right ) \right\rbrace,\\
\Esub{\log p(\boldsymbol{\theta} | \boldsymbol{\alpha}^1,  \boldsymbol{\alpha}^2) }{\boldsymbol{\theta}} &=& \sum^G_{g=1} \sum^G_{h=1} \log \Gamma \left(\alpha^1_{gh} + \alpha^2_{gh} \right ) - \log \Gamma (\alpha^1_{gh}) - \log \Gamma (\alpha^2_{gh}) \\ 
 &+& \sum^G_{g=1} (\alpha^1_{gh} -1) \left \lbrace \Psi(\zeta^1_{gh}) - \Psi(\zeta^1_{gh}+\zeta^2_{gh}) \right\rbrace \\
 &+&  \sum^G_{g=1} (\alpha2_{gh} -1) \left \lbrace \Psi(\zeta^2_{gh}) - \Psi(\zeta^1_{gh}+\zeta^2_{gh}) \right\rbrace .\\
 \end{eqnarray*}
 The terms for the second part of the lower bound are given below:
 \begin{eqnarray*}
 \Esub{\log q(\mathbf{Z}_{ij}^1 | \boldsymbol{\phi}_{ij}^1)}{\mathbf{Z}^1} &=& \Esub{ \sum^G_{g=1} {Z}_{ijg}^1  \log{\phi}_{ijg}^1}{\mathbf{Z}^1}\\
&=& \sum^G_{g=1} \phi_{ijg}^1  \log{\phi}_{ijg}^1, \\
\Esub{\log q(\mathbf{Z}_{ij}^2 | \boldsymbol{\phi}_{ij}^2)}{\mathbf{Z}^2} &=& \sum^G_{g=1} \phi_{ijg}^2  \log{\phi}_{ijg}^2, \\
\Esub{\log q(\boldsymbol{\tau_n} | \boldsymbol{\gamma_n})}{\boldsymbol{\tau}} &=& \Esub{\log \Gamma \left(\sum^G_{h=1} \gamma_{nh} \right) - \sum^G_{k=1} \log \Gamma (\gamma_{nk}) + \sum^G_{g=1} (\gamma_{ng} -1) \log \tau_{ng}}{\boldsymbol{\tau}} \\
&=& \log \Gamma \left(\sum^G_{h=1} \gamma_{nh} \right) - \sum^G_{k=1} \log \Gamma (\gamma_{nk}) + \sum^G_{g=1} (\gamma_{ng} -1) \times \left \lbrace \Psi({\gamma}_{ng}) - \Psi\left( \sum^G_{k=1} \gamma_{nk} \right ) \right\rbrace,\\
 \Esub{\log q(\boldsymbol{\theta} | \boldsymbol{\zeta}^1, \boldsymbol{\zeta}^2 )}{\boldsymbol{\theta}} &=& \sum^G_{g=1} \sum^G_{h=1} \log \Gamma \left(\zeta^1_{gh} + \zeta^2_{gh} \right ) - \log \Gamma (\zeta^1_{gh}) - \log \Gamma (\zeta^2_{gh}) \\ 
 &+& \sum^G_{g=1} (\zeta^1_{gh} -1) \left \lbrace \Psi(\zeta^1_{gh}) - \Psi(\zeta^1_{gh}+\zeta^2_{gh}) \right\rbrace \\
 &+&  \sum^G_{g=1} (\zeta^2_{gh} -1) \left \lbrace \Psi(\zeta^2_{gh}) - \Psi(\zeta^1_{gh}+\zeta^2_{gh}) \right\rbrace.
\end{eqnarray*}
To estimate $\boldsymbol{\hat{\beta}}$ we make us of a Newton-Raphson step to iteratively maximise ${\mathcal{L}}.$ It's simpler to first calculate the gradient and Hessian functions in terms of $\boldsymbol{\delta}$: 
\begin{eqnarray*}
\del{\delta_i }{\cal L} &=& \del{\delta_i}{~} \sum^N_{n=1} \Esub{\log p (\tau_n | \delta)}{\tau} \\ 
 &=& \del{\delta_i }{~} \sum^N_{n=1} \left [ \log \Gamma \left ( \sum^G_{h=1} \delta_{h} \right )  - \sum^G_{h=1} \log \Gamma (\delta_{h}) + \sum^G_{h=1} (\delta_{h}-1)\left \{ \Psi (\gamma_{ng}) - \Psi(\sum^G_{h=1} \gamma_{nh} )\right \} \right ]\\
 &=& \del{\delta_i }{~} N \left \{  \log \Gamma \left ( \sum^G_{h=1} \delta_h \right )  -  \sum^G_{k=1} \log \Gamma (\delta_k)  \right \} + \del{\delta_i }{~} \sum^N_{n=1} \sum^G_{g=1} (\delta_g -1)\left \{ \Psi (\gamma_{ng}) - \Psi(\sum^G_{h=1} \gamma_{nh} )\right \} \\
 &=& N \left \{  \Psi \left ( \sum^G_{h=1} \delta_h \right )  -  \Psi (\delta_i)  \right \} + \sum^N_{n=1} \left \{ \Psi (\gamma_{ni}) - \Psi(\sum^G_{h=1} \gamma_{nh} )\right \} .\\
\Rightarrow \deltwo{\delta_i}{ \delta_j}{\cal L} &=& N \left \{  \Psi^{'} \left ( \sum^G_{h=1} \delta_h \right )   -   {\mathbb I}_{i=j}\Psi^{'}\left (\delta_i \right ) \right \}.
\end{eqnarray*}

Now, noting that $\del{\beta_{gp} }{\delta_{ig}} = W_{ip}\exp(\sum^P_{p=1} W_{ip}\beta_{gp}), $
we can then maximise the lower bound ${\mathcal{L}}$ with respect to  $\boldsymbol{\beta}$ by making use of the chain rule: 

\begin{eqnarray*}
\del{\beta_{iq} }{\mathcal{L}} &=& \sum^N_{n=1}  \del{\delta_{ni} }{\mathcal{L}} \del{\beta_{iq} }{\delta_{ni}} \\
&=& \sum^N_{n=1} W_{nq}\exp(\sum^P_{p=1} W_{np}\beta_{ip}) \\
& \times & \left \{ \Psi \left [ \sum^G_{h=1}  \exp(\sum^P_{p=1} W_{np}\beta_{hp}) \right ]  - \Psi \left [ \exp(\sum^P_{p=1} W_{np}\beta_{ip}) \right] + \Psi (\gamma_{ng}) - \Psi(\sum^G_{h=1} \gamma_{nh} )\right \}.
\end{eqnarray*}

We can then calculate the Hessian matrix, again making use of the product rule:

\begin{eqnarray*} \deltwo{\beta_{iq}}{ \beta_{jr}}{\cal L} &=& \sum^N_{n=1} W_{nq}\exp(\sum^P_{p=1} W_{np}\beta_{ip}) \\
& \times & \del{\beta_{jr} }{~} \left \{ \Psi \left [ \sum^G_{h=1}  \exp(\sum^P_{p=1} W_{np}\beta_{hp}) \right ]  -  \Psi \left [ \exp(\sum^P_{p=1} W_{np}\beta_{ip}) \right] + \Psi (\gamma_{ni}) - \Psi(\sum^G_{h=1} \gamma_{nh} )\right \} \\
&+& \sum^N_{n=1} \del{\beta_{jr} }{~} \left( W_{nq}\exp(\sum^P_{p=1} W_{np}\beta_{ip}) \right ) \\
& \times & \left \{ \Psi \left [ \sum^G_{h=1}  \exp(\sum^P_{p=1} W_{np}\beta_{hp}) \right ]  -  \Psi \left [ \exp(\sum^P_{p=1} W_{np}\beta_{ip}) \right] + \Psi (\gamma_{ni}) - \Psi(\sum^G_{h=1} \gamma_{nh} )\right \}. \\
\\
&=& \sum^N_{n=1} W_{nq}\exp(\sum^P_{p=1} W_{np}\beta_{ip}) \\
&\times & W_{nr}\exp(\sum^P_{p=1} W_{np}\beta_{jp})\left \{ \Psi^{'} \left [ \sum^G_{h=1}  \exp(\sum^P_{p=1} W_{np}\beta_{hp}) \right ]  -  {\mathbb I}_{i=j}\Psi^{'} \left [ \exp(\sum^P_{p=1} W_{np}\beta_{ip}) \right] \right \}\\
&+& {\mathbb I}_{i=j} \left( W_{nq} W_{nr}\exp(\sum^P_{p=1} W_{np}\beta_{ip}) \right. \\
& \times & \left . \left \{ \Psi \left [ \sum^G_{h=1}  \exp(\sum^P_{p=1} W_{np}\beta_{hp}) \right ]  -  \Psi \left [ \exp(\sum^P_{p=1} W_{np}\beta_{ip}) \right] + \Psi (\gamma_{ni}) - \Psi(\sum^G_{h=1} \gamma_{nh} )\right \} \right ) . \\
\end{eqnarray*}

Cleaning this up a little gives the result:
\begin{eqnarray*}
 \deltwo{\beta_{iq}}{ \beta_{jr}}{\cal L} &=& \sum^N_{n=1} W_{nq} W_{nr} \exp(\sum^P_{p=1} W_{np}(\beta_{ip} + \beta_{jp})) \left \{ \Psi^{'} \left [ \sum^G_{h=1}  \exp(\sum^P_{p=1} W_{np}\beta_{hp}) \right ]  -  {\mathbb I}_{i=j}\Psi^{'} \left [ \exp(\sum^P_{p=1} W_{np}\beta_{ip}) \right] \right \} \\
&+& {\mathbb I}_{i=j} \left ( W_{nq} W_{nr}\exp(\sum^P_{p=1} W_{np}\beta_{ip}) \right .\\
& \times & \left . \left \{ \Psi \left [ \sum^G_{h=1}  \exp(\sum^P_{p=1} W_{np}\beta_{hp}) \right ]  -  \Psi \left [ \exp(\sum^P_{p=1} W_{np}\beta_{ip}) \right] +   \Psi (\gamma_{ni}) - \Psi(\sum^G_{h=1} \gamma_{nh} )\right \} \right ).
\end{eqnarray*}

\end{document}